\pdfoutput=1
\documentclass[UKenglish,texlive=2011,cernpreprint,txfonts]{latex/atlasdoc}

\usepackage[biblatex=false]{latex/atlaspackage}
\usepackage{latex/atlasphysics}
\usepackage{cite}
\usepackage{url}
\usepackage{hyperref}

\graphicspath{{logos/}}

\usepackage{chngcntr}
\counterwithout{equation}{section} 

\AtlasTitle{\boldmath  A search for prompt lepton-jets in $pp$ collisions at $\sqrt{s}=8$\,TeV with the ATLAS detector}
\PreprintIdNumber{CERN-PH-EP-2015-242}
\AtlasJournal{JHEP}

\AtlasAbstract{A search is presented for a new, light boson with a mass of about 1 GeV and decaying promptly to jets of collimated electrons and/or muons (lepton-jets). The analysis is performed with 20.3 fb$^{-1}$ of data
  collected by the ATLAS detector at the Large Hadron Collider in proton--proton collisions at a centre-of-mass energy of 8 TeV. Events are required to contain at least two
  lepton-jets. This study finds no statistically significant deviation from predictions of the Standard Model and places 95\% confidence-level upper limits on the contribution of new phenomena beyond the SM, incuding SUSY-portal and Higgs-portal models, on the number of events with lepton-jets.}

\AtlasCoverCommentsDeadline{23 November 2015}

\AtlasCoverSupportingNote{ATL-COM-PHYS-2014-454}{https://cds.cern.ch/record/1701078}

\AtlasCoverAnalysisTeam{Mahsana Haleem, Joseph M. Izen,  Bernhard Meirose(*), Harisankar Namasivayam}

\AtlasCoverEdBoardMember{Claire Adam Bourdarios (*), Antonio Boveia, Else Lytken, Jonas Strandberg}

\AtlasCoverEgroupEditors{atlas-exot-2014-09-editors@cern.ch}

\AtlasCoverEgroupEdBoard{atlas-exot-2014-09-editorial-board@cern.ch}

\newcommand{\intlumi}{20.3~fb$^{-1}$\xspace}

\newcommand{\fb}{$\rm fb^{-1}$\xspace}

\newcommand{\LJ}{LJ\xspace}

\newcommand{\gammad}{$\gamma_{\rm d}$\xspace}

\def\gammad{$\gamma_d$}
\def\mgammad{$m_{\gamma_{d}}$}

\def\ngammad{$n_{\gamma_{d}}$\xspace}

\begin{document}

\title{\boldmath  A search for prompt lepton-jets in $pp$ collisions at $\sqrt{s}=8$\,TeV with the ATLAS detector}
\author{The ATLAS Collaboration}


\maketitle

%

%

%

%

\section{Introduction}
\label{sec:intro}
\noindent In several models of physics beyond the Standard Model (SM) \cite{ArkaniHamed1,Yavin1,Yavin2,falkowski,HiggsToDark2}, the so-called dark matter (see e.g. Ref.~\cite{Bi} and references therein) is charged under a non-Abelian, dark-sector, gauge symmetry that is broken at an energy scale $\mathcal{O}(1~\text{GeV})$. The dark-sector ground state can transition to and from excited states via the emission of a dark gauge boson, referred to as the dark photon ($\gamma_d$),  that couples very weakly to the SM particles via kinetic mixing~\cite{Holdom} with the SM photon. In these models, the Large Hadron Collider (LHC) could produce excited dark-sector states via their interactions with particles found in models of supersymmetry (SUSY)~\cite{ArkaniHamed1,Yavin2} or with Higgs scalar bosons \cite{falkowski,HiggsToDark2} (here referred to as SUSY-portal and Higgs-portal models, respectively), which then decay via the emission of dark photons. If dark photons carry masses of $\mathcal{O}(1~\text{GeV})$, then the dark photon produced from the decay chain of heavier particles such as the SM Higgs boson or SUSY particles would be highly boosted. Depending on its mass, the dark photon would decay primarily into a collimated pair of leptons or light hadrons. The leptonic final-state is experimentally easier accessible, offering a distinct signature that stands out amongst large hadronic backgrounds. A collimated set of energetic leptons is referred to as a lepton-jet (LJ).\\

\noindent A search is carried out for final-states with two prompt lepton-jets using data accumulated in proton--proton collisions at a centre-of-mass energy $\rts = 8$~TeV with the ATLAS detector~\cite{atlas}. Many new physics models predict at least two lepton-jets in the final-states as described in Refs.~\cite{Yavin2,falkowski}. The analysis focuses on the presence of lepton-jets and does not rely on the rest of the event topology. The dark-photon decay width, $\Gamma_\ell$, and the kinetic mixing parameter, $\epsilon$, are related through

\begin{eqnarray}
\Gamma_\ell = \frac{1}{3}\alpha \epsilon^2 m_{\gamma_d} \sqrt{1-\frac{4 m_\ell^2}{m_{\gamma_d}^2}} \left (1+ \frac{2 m_\ell^2}{m_{\gamma_d}^2} \right)  ,
\label{eqn:gamma}
\end{eqnarray}     

\noindent where $\alpha$ is the fine structure constant and $m_{\gamma_d}$ and $m_\ell$ denote the masses of the dark photons and charged leptons, respectively \cite{DecayLengthEps, MSUSYHV}. The analysis focuses on dark photons with prompt-decays, i.e. consistent with zero decay length within the experimental resolution. Previous searches for prompt lepton-jets, with ATLAS data at $\sqrt{s}=7$\,TeV, resulted in upper limits on the production of two lepton-jets in a SUSY-portal model~\cite{SUSYPromptLJ7TeVATLAS} and for a Higgs-portal model~\cite{HiggsPromptLJ7TeVATLAS}. The CMS and D0 collaborations also set upper limits on prompt lepton-jet production~\cite{cms_lj1,cms_lj2,CMS_muons, d0_lj1, d0_lj2}. Related searches for non-prompt lepton-jets \cite{HiggsDisplacedLJ8TeVATLASpaper} have been performed by ATLAS and have set constraints on smaller values of the kinetic mixing parameter, $\epsilon \le 10^{-5}$. There are additional constraints on the kinetic mixing parameter and dark-photon mass, e.g. from beam-dump and fixed target experiments~\cite{DecayLengthEps,beam-dump1,beam-dump2,beam-dump3,fixed-target1,fixed-target2,fixed-target3,fixed-target4,fixed-target5,dg-lowE}, $e^{+}e^{-}$ collider experiments~\cite{epluseminus-colliders1, Babusci1, Babusci2, Anastasi, epluseminus-colliders2,epluseminus-colliders4}, electron and muon magnetic moment measurements~\cite{emu-magneticmoment1,emu-magneticmoment2} and astrophysical observations~\cite{astrophysexp1,astrophysexp2}.

\section{The ATLAS detector}
\label{sec:ATLAS}

\noindent  ATLAS is a multi-purpose detector~\cite{atlas} consisting of an
inner tracking detector (ID), electromagnetic and hadronic calorimeters and a muon spectrometer (MS)
that employs toroidal magnets. The ID provides precision tracking of charged
particles for pseudorapidity\footnote{ATLAS uses a right-handed coordinate system with its origin at the nominal
 interaction point (IP) in the centre of the detector. The
  $z$-axis points along the beam pipe. The $x$-axis points from the IP to the
  centre of the LHC ring, and the $y$ axis points upward. Cylindrical
  coordinates $(r,\phi)$ are used in the transverse plane, $\phi$
  being the azimuthal angle around the beam pipe. Pseudorapidity
  is defined in terms of the polar angle $\theta$ as
  $\eta=-\ln\tan(\theta/2)$.} $|\eta|<2.5$ using silicon pixel and silicon
microstrip (SCT) detectors and a straw-tube transition radiation tracker (TRT)
that relies on transition radiation to distinguish electrons from
pions in the range $|\eta| < 2.0$. 

\noindent  The sensors of the pixel detector have a typical pixel size of $50 \times 400$  $\si{\micro}$m and typically provide three spatial measurements along the track of a charged particle. The innermost layer with a radial distance to the beamline of about 5 cm is known as the B-layer. The SCT has sensors with a strip pitch of 80 $\mu$m and provides eight measurements for a typical track. The fine-grained sensors of the semiconductor trackers permit the reconstruction of the closely aligned tracks of lepton-jet candidates (Section \ref{sec:ljdefinition}).

\noindent  The liquid-argon (LAr) electromagnetic (EM) sampling calorimeters cover the range $|\eta| <
3.2$. The calorimeter's transverse granularity, typically $\Delta\eta \times \Delta\phi$ of
$0.025 \times 0.025$, and three-fold shower-depth segmentation are used to construct discriminating variables for evaluating the electromagnetic character of lepton-jet candidates (Section \ref{sec:LJbkg}).

\noindent  A scintillator-tile calorimeter, divided into a barrel and two extended-barrel cylinders,
on each side of the central barrel, provides hadronic calorimetry
in the range $|\eta| < 1.7$, while a LAr hadronic end-cap calorimeter 
provides coverage over $1.5<|\eta|<3.2$. The LAr forward calorimeters provide both, electromagnetic and hadronic energy
measurements, and extend the coverage to $|\eta| \le 4.9$. The
calorimeter system has a minimum depth of 9.7 nuclear interaction lengths at
$\eta = 0$. The MS is a large tracking system, consisting of three parts: a magnetic field provided by three toroidal magnets, a set of 1200 chambers measuring with high spatial precision the tracks of the outgoing muons, a set of triggering chambers with accurate time-resolution. It covers $|\eta|<2.7$ and provides precision tracking and triggering for muons.

\noindent  ATLAS has a three-level trigger system. The Level 1 (L1) trigger is implemented in hardware, and it uses information from the calorimeters and muon spectrometer to reduce the event rate
to 75--100\,kHz. The software-based Level 2 (L2) trigger and the Event Filter (EF) reduce the event rate to 300--500~Hz of events that are retained for offline analysis.
The L1 trigger generates a list of region-of-interest (RoI) $\eta$--$\phi$ coordinates. The muon RoIs have a spatial extent of 0.2 in $\Delta \eta$ and $\Delta \phi$ in the MS barrel, and 0.1 in the MS
endcap. Electromagnetic calorimeter RoIs have a spatial extent of 0.2 in $\Delta \eta$ and $\Delta \phi$. For the L2 trigger the reconstruction is mostly based on simplified algorithms running on data localized in the RoI which was reported by L1. At the EF level the trigger system has access to the full event for processing.

%
\section{Signal models}
\label{sec:signalsamples}
Two benchmark models are used to interpret the data. In the SUSY-portal model (Section~\ref{sec:signalSUSYmodel}),
a pair of squarks is produced and the cascade decays of the squarks include dark-sector particles and one or more dark photons. In the 
Higgs-portal model (Section~\ref{sec:signalHiggsmodel}), the SM Higgs boson decays into a pair of dark fermions, each of which decays into one or more dark photons in cascades. For both models, the dark photons decay into lepton pairs, that can be reconstructed as a lepton-jet, or light hadrons, depending on the branching fractions. Monte Carlo (MC) simulated samples are produced for the two models. All signal MC events are processed with the \textsc{Geant4}-based ATLAS detector simulation~\cite{Geant1,atlas_simulation} and then analysed with the standard ATLAS reconstruction software. The branching ratio (BR) values for the dark-photon decays to leptons are taken from Ref.~\cite{falkowski}. In all signal models used to interpret the results the dark photons are required to decay promptly with mean life time  ($c \tau$) close to zero. For the Higgs-portal model, long-lived dark photon samples with $c\tau = 47$ mm are used to extrapolate the signal efficiency of zero $c\tau$ dark photons to non-zero $c\tau$ dark photons (Section \ref{sec:limits}). \\

 \subsection{SUSY-portal lepton-jet MC simulation}
 \label{sec:signalSUSYmodel}

\begin{figure}[ht!]
\includegraphics[width=0.45\textwidth]{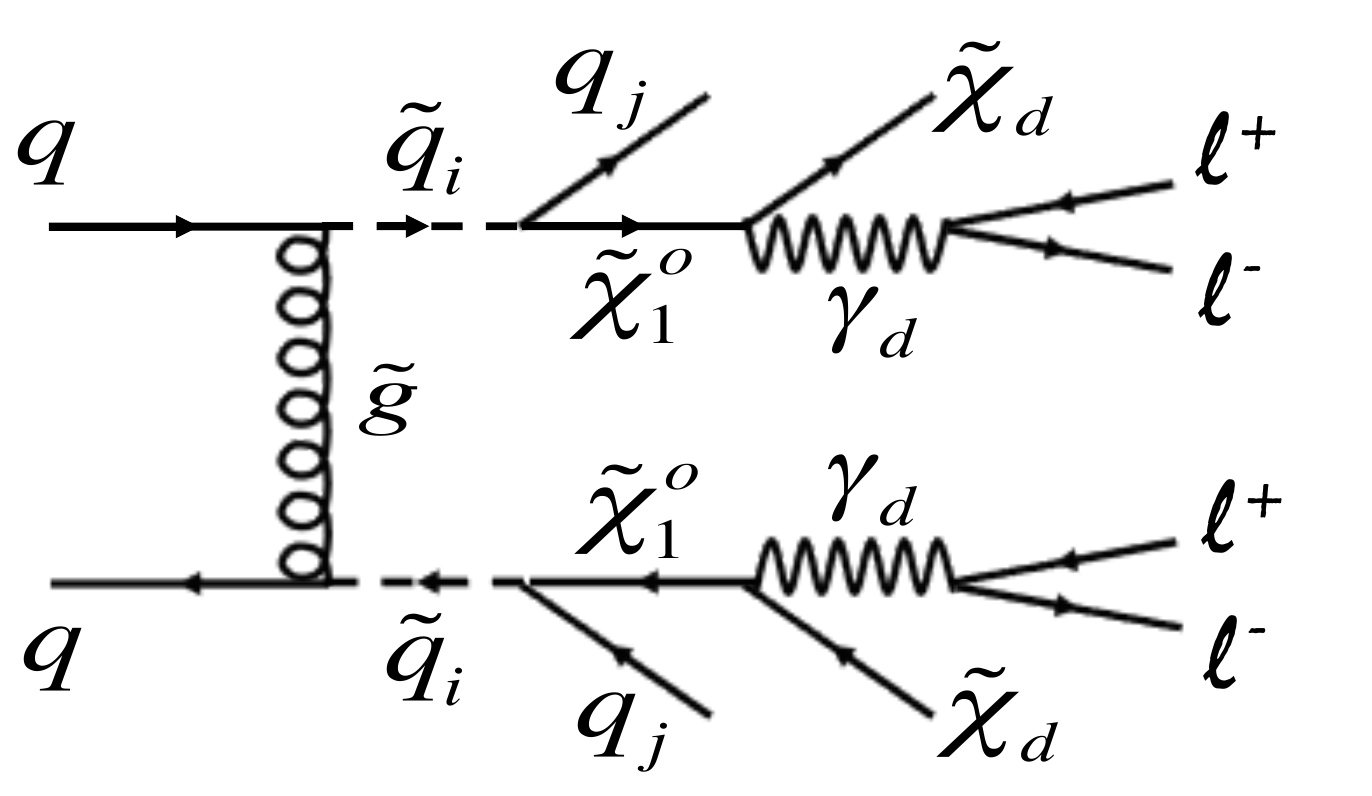}
\includegraphics[width=0.5\textwidth]{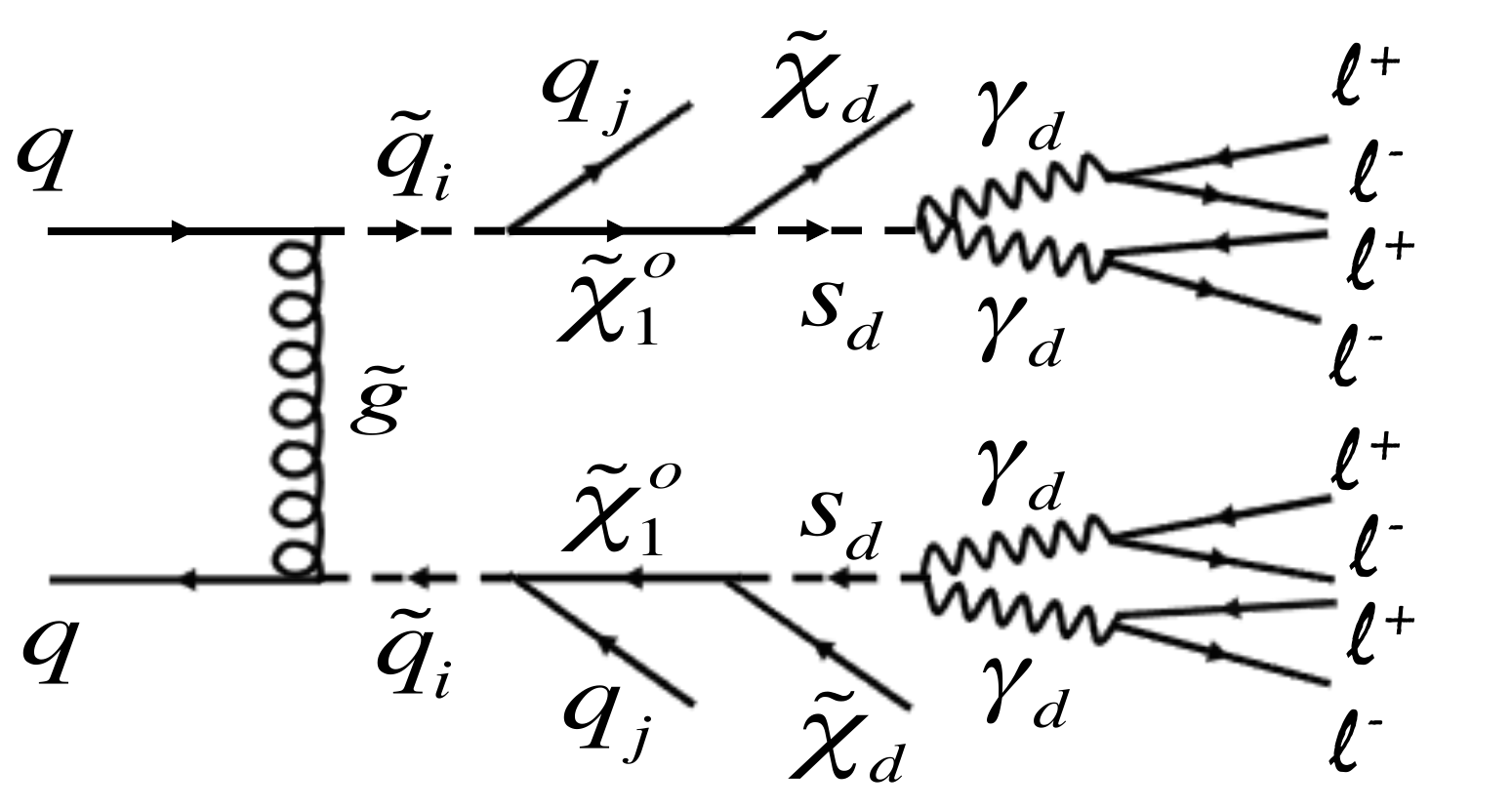}
\caption{Feynman diagram illustrating the dark-photon production in the $2 \gamma_d$ final-state (left), and $4 \gamma_d$ final-state (right).} 
\label{fig:feydiagram1}
\end{figure}

\noindent A benchmark SUSY model~\cite{Yavin2}  is used to simulate SUSY production of dark-sector particles and dark photons. Simulated samples are produced in several steps. Squark ($\tilde{q}$) pair events are generated with \textsc{Madgraph}~\cite{Madgraph}, version 5, in a simplified model with light-flavour squark pairs with decoupled gluinos~\cite{Alwall2008, Alves2011}.\footnote{This is the same simplified model used in a previous ATLAS search and shown in the third plot of Figure 10 in Ref. \cite{Aad2014wea}. In the analysis context, the fact that gluinos are decoupled implies the $2 \rightarrow 2$ production, such that there are two SUSY particles at the hard scatter producing two lepton-jets per event.}. Then \textsc{Bridge}~\cite{BRIDGE}, interfaced with \textsc{Madgraph}, simulates squark decays into neutralinos. The neutralinos decay into dark-sector particles, which decay to SM particles as shown in Figure \ref{fig:feydiagram1}. The squarks are set to decay with a $100\%$ BR into a quark and a neutralino ($\tilde{\chi}_1^{0}$). The neutralinos decay into dark-sector particles in two ways: $\tilde{\chi}_1^{0} \rightarrow \gamma_d \tilde{\chi}_d$ or $\tilde{\chi}_1^{0} \rightarrow s_d \tilde{\chi}_d$, where $s_d$ is a dark scalar particle that decays to $\gamma_d \gamma_d$ and $\tilde{\chi}_d$ is a dark neutralino. In this model, the stable dark-matter particle is the dark neutralino which is invisible in the detector.  For fragmentation and hadronization \textsc{Pythia} 8 \cite{Pythia, Pythia8} is used, with the CTEQ6L1 1~\cite{CTEQ6} PDF parton distribution function (PDF) set, and the AUET2  \cite{mc11tune} set of tuned parameters.\\ 
 
\noindent  As the dark-sector is loosely constrained experimentally, the squark mass, the dark-photon mass, and all intermediate masses are chosen to correspond to well-motivated nominal values and ranges to which the search is sensitive. The squark-pair production cross section, and hence the signal model sensitivity, would decrease with an increase in the squark mass. The squark mass also affects the sensitivity through the boost of the final-state dark photon. The squark mass,  $m_{\tilde{q}}$, is considered to be 700 GeV, which is motivated by the upper limit (17 fb) on the cross section times BR established by a previous search for prompt lepton-jets \cite{SUSYPromptLJ7TeVATLAS} at  $\rts = 7$~TeV.  This translates into an upper limit of 77 fb on the squark pair-production cross section, which is the predicted cross section at a squark mass of nearly 700 GeV. \\

\noindent  For the dark-sector, a $\sim$1 GeV dark-photon mass is considered, as predicted by the SUSY-portal and Higgs-portal models. For consistency with the $\mathcal{O}(1~\text{GeV})$ scale for dark-sector particle masses, following the model described in Ref.~\cite{Yavin2}, the set of mass values for the squark decay products is chosen as follows: the $\tilde{\chi}_1^{0}$ mass is set to $8$ GeV, and a set of $\gamma_d$ mass values is chosen between $0.1$ GeV and $2$ GeV. A dark photon that kinetically mixes with a SM photon would couple with the same strength to lepton--antilepton and $q\bar{q}$ pairs as the SM photon. The virtual photon conversion rates are measured in low-energy $e^+ e^-$ annihilations ~\cite{PDG2012,SLACloweeannihilation} and a few sets of $\gamma_d$ mass points are chosen for which the BRs into the leptons are large enough (e.g. 0.25--1.0) to be detectable. The BRs to leptons are in the range from 25\% to 100\% except when $m_{\gamma_d}$ is close to either the $\rho$ or $\phi$ mass. Table \ref{tab:massvalues} summarizes the particle masses assumed.

\begin{table}[ht!]
\centering
\caption{Mass points for SUSY and dark-sector particles in GeV.}
\begin{tabular}{| c | c | c | c | c |}
\hline 
 $m_{\tilde{q}} $&   $m_{\tilde{\chi}_1^{0}}$   &   $m_{\tilde{\chi}_d}$    &  $m_{\gamma_d}$ &   $m_{s_d}$ \\
\hline
 700	&  8	& 2 &  0.1, 0.2, 0.3, 0.4, 0.5, 0.7, 0.9, 1.2, 1.5, 2.0  & 2, 4, 4.5 \\
\hline
\end{tabular}
\label{tab:massvalues}
\end{table}

\noindent The intermediate dark-sector masses are chosen such that all  particles remain on-shell. The dark scalar $s_d$ mass is set to $2$ GeV for samples with $\gamma_d$ masses below $0.9$ GeV, and it is set to $4$ GeV for samples with $\gamma_d$ mass in the range from $0.9$ to $1.5$ GeV. A  $4.5$ GeV $s_d$ mass is used for the MC sample with a $\gamma_d$ mass of $2$ GeV.  The model sensitivity is mostly driven by the dark photons' boost and BR to leptons.

%
 \subsection{Higgs-portal lepton-jet MC simulation}
 \label{sec:signalHiggsmodel}

\begin{figure}[ht!]
		\includegraphics[width=0.5\textwidth]{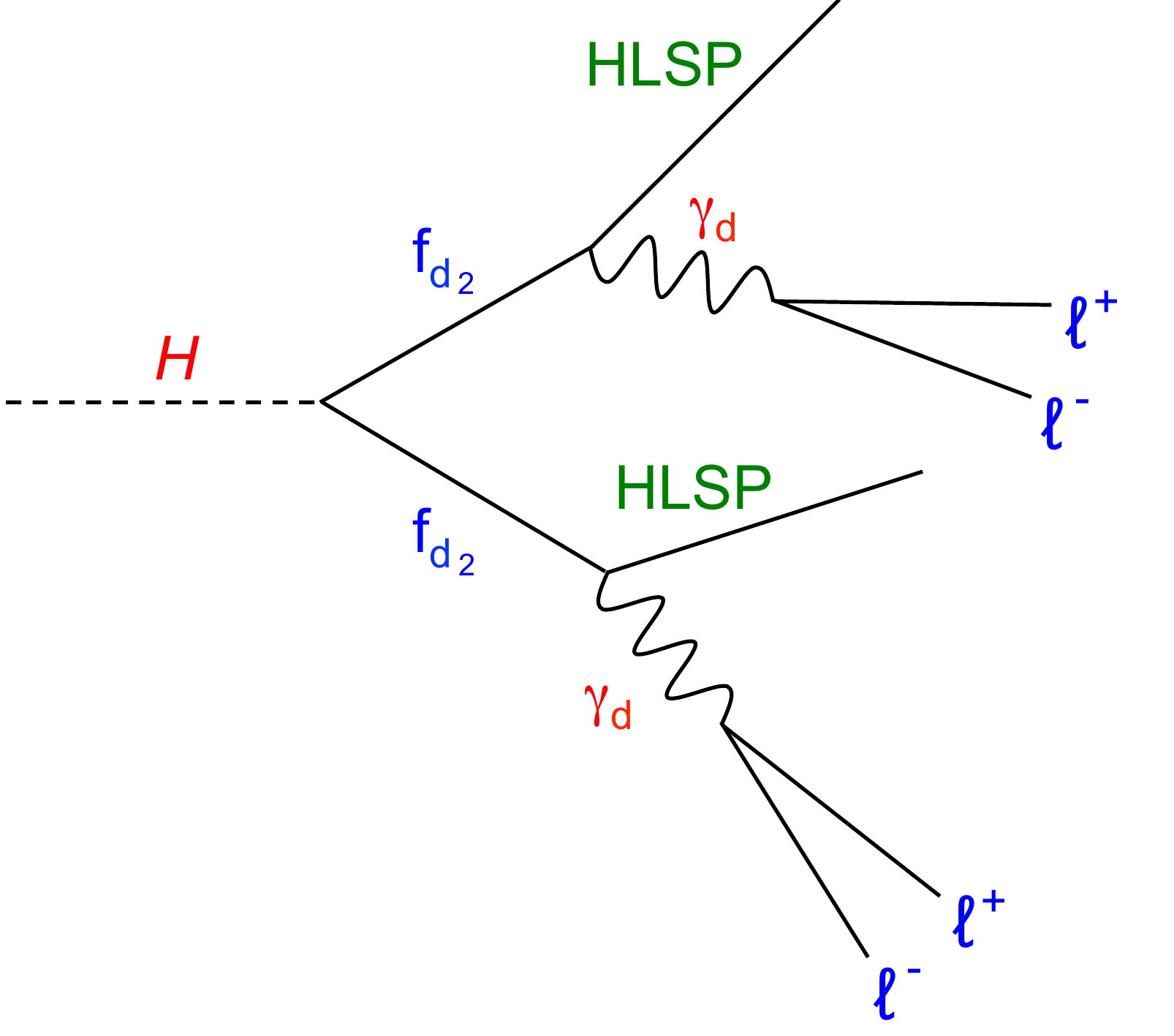} 
		\label{fig:frvz2gd}
	        \includegraphics[width=0.5\textwidth]{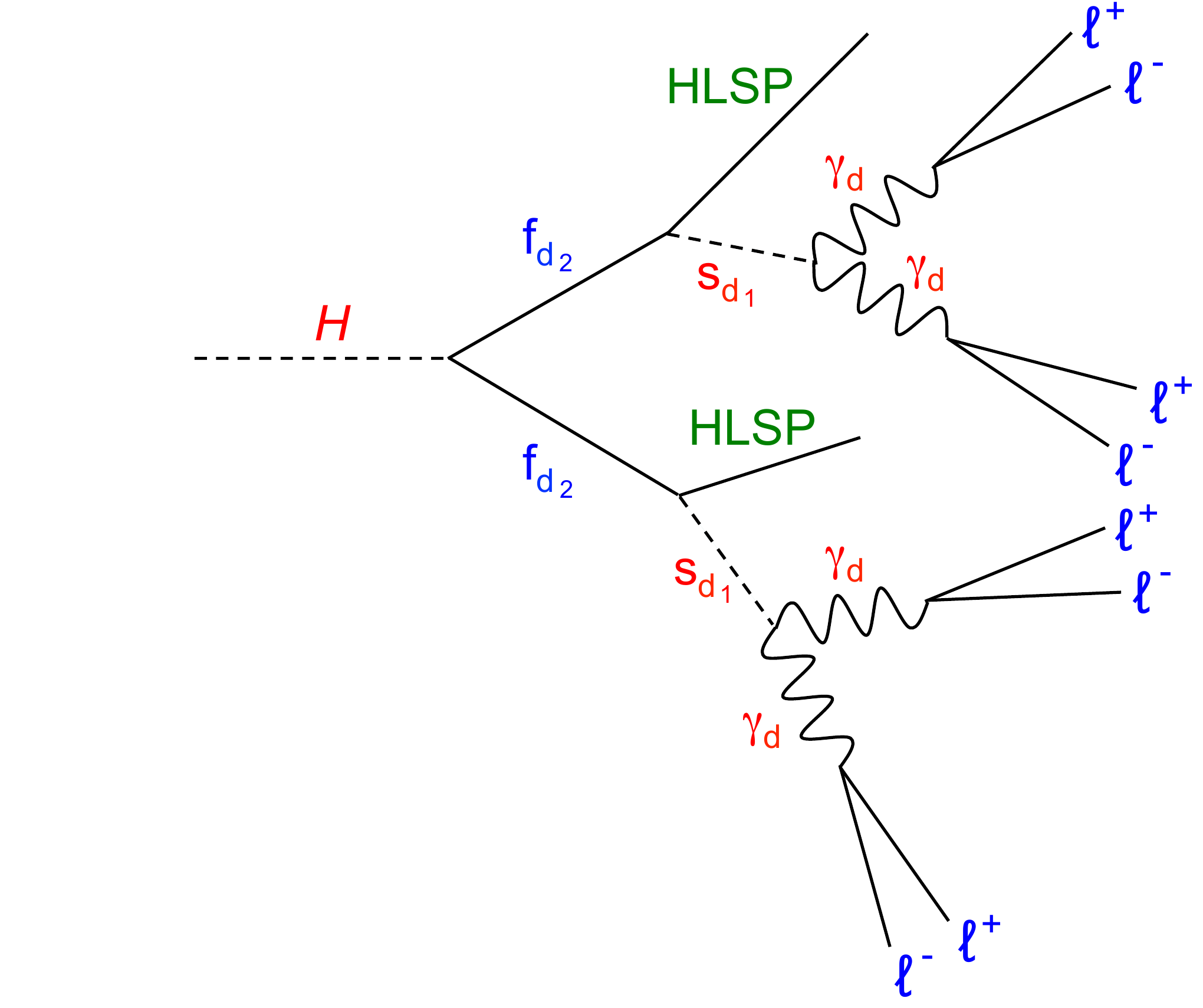} 
		\label{fig:frvz4gd}
	\caption{The Higgs boson decays to a pair of dark fermions $f_{d_2}$, each of which decays to a Hidden Lightest Stable Particle (HLSP) and a dark photon (left) or to a HLSP and a dark scalar $s_{d_1}$ (right) that in turn decays to a pair of dark photons \gammad.} 
	\label{fig:FRVZ_model}
\end{figure}

A hypothesized decay of the Higgs boson to a pair of dark fermions $f_{d_2}$ is considered \cite{falkowski, HiggsToDark2} as shown in  Figure \ref{fig:FRVZ_model}. Dark fermions $f_{d_2}$ decay to a dark photon ($\gamma_d$) and a lighter dark fermion ($f_{d_1}$) or the Hidden Lightest Stable Particle (HLSP) (Figure \ref{fig:FRVZ_model}, left). In another process, a dark fermion $f_{d_2}$ decays to a lighter dark fermion $f_{d_1}$ (referred to as HLSP) and a dark scalar $s_{d_1}$. The $s_{d_1}$ decays to a pair of dark photons (Figure \ref{fig:FRVZ_model}, right). The same set of generators (\textsc{Madgraph} +  \textsc{Bridge} +  \textsc{Pythia} 8 chain) that are used to generate the SUSY samples are used to generate the Higgs-portal samples. Also the same PDF set and underlying event tune is used.\\

\noindent The Higgs boson ($m_{H} = 125$ GeV) is generated through the gluon-fusion production mechanism with an estimated production cross section of $\sigma_{ \text{SM}}$ = 19.2 pb for $pp$ collisions at $\sqrt{s} = $ 8 TeV \cite{HiggsCross}. The mass of $f_{d_2}$ is chosen to be 5 GeV, $f_{d_1}$ and $s_{d1}$ masses are chosen to be 2 GeV, and the dark photon ($\gamma_d$) mass is chosen to be 0.4 GeV. For consistency, the choice of Higgs boson mass and the dark-sector particles' masses in the cascade decay is the same as used in the ATLAS displaced lepton-jets analysis \cite{HiggsDisplacedLJ8TeVATLASpaper}. 

\section{Pre-selection of events}
\label{sec:eventsel}

\noindent  Events are required to have a primary collision vertex containing at least three tracks with transverse momentum \pt > 400 MeV. All events must satisfy the trigger, and offline reconstructed objects (electrons or muons) are required to match the leptons firing the trigger.

Unprescaled triggers with the lowest available trigger threshold are used, and a logical OR of triggers is taken to maximize the signal acceptance. For the electron channels, a single-electron trigger with a transverse energy threshold of 60 GeV as well as a trigger requiring two electromagnetic showers with minimum transverse energies of $\et > 35$ GeV and $\et > 25$ GeV are used. For the muon channels, a dimuon trigger with a \pt threshold of 13 GeV as well as a single-muon trigger with a \pt threshold of 36 GeV are used. For the mixed channels where both electrons and muons are present, the single-electron, the single-muon and the dimuon triggers are used. \\

\noindent Electron candidates to be used to build lepton-jets are reconstructed from clusters of deposited energy with $\et > 10$ GeV inside the EM calorimeter fiducial region, $|\eta| < 2.47$, excluding the barrel/end-cap transition region $1.37 < |\eta| < 1.52$ where there is substantial inactive material that is difficult to model accurately. Each cluster must have at least one inner detector track associated. The reconstructed electron is required to match an electron trigger object above the \et\ trigger threshold in the trigger system within $\Delta R \equiv \sqrt{(\Delta \phi)^2 + (\Delta \eta)^2} < 0.2$. The transverse shower profiles of these reconstructed electrons differ with respect to an isolated electron from a $W$ or $Z$ boson because the electrons overlap.\\

\noindent Muon candidates to be used to build lepton-jets must be reconstructed in both the ID and the MS and have $|\eta| < 2.5$. Additional requirements are placed on the number of associated hits in the silicon pixel and microstrip
detectors, as well as on the number of track segments in the MS. A requirement $|d_0| < 1$ mm with respect to the primary vertex is imposed on muons. Muon candidates are required to match to the muon trigger objects within $\Delta R < 0.2$.

\subsection{Track selection}
\label{sec:TrackandObjsel}

\noindent The track selection criteria are crucial for reconstruction of close pairs of tracks and for assessment of fake rates (e.g. when a single track is misreconstructed as two tracks). The criteria are as follows:
		\begin{itemize}
			\item \pt \textgreater~ 5 GeV, \textbar $\eta$ \textbar~ \textless~ 2.5.
			\item Transverse impact parameter $|d_0| < 1$ mm.
			\item Number of B-layer hits $\geq$ 1.
			\item Number of Pixel-layer hits $\geq$ 2 (includes the B-layer hit requirement).
			\item Number of Pixel + SCT-layer hits $\geq$ 7.
                                \item Longitudinal impact parameter $|z_0 \sin \theta| < 1.5$ mm.
		\end{itemize}

All tracks are required to come from the same primary vertex.

\section{Selection of lepton-jets}
\label{sec::LJselec}
%

Signal MC events together with background MC events and background-dominated data from a jet-triggered sample are used to develop optimized criteria that are applied to pre-selected events to preferentially retain \LJ events while rejecting backgrounds (Section~\ref{sec:LJbkg}).
A data-driven method is used to determine the background content in the final sample of \LJ candidate events  (Section~\ref{sec:bkg_est}).

\subsection{Lepton-jet definition}
\label{sec:ljdefinition}
Lepton-jets are bundles of tightly collimated, high-$\pt$ leptons. In the current study,  only prompt $\gamma_d$ leptonic decays ($e^+ e^-$ or $\mu^+ \mu^-$) are selected. Hadronic $\gamma_d$ decays cannot be distinguished from multjiet background. \footnote{The simulation does include the hadronic decays. However, no requirements are applied to select the hadronic decay products, as such selection would be masked by the multijet background.} Non-prompt-decays suffer lower multijet backgrounds and are treated elsewhere \cite{HiggsDisplacedLJ8TeVATLASpaper}. \\

\noindent  Two prompt-decay scenarios are considered. In the first, a single $\gamma_d$ decays into $e^+ e^-$, $\mu^+ \mu^-$ or $\pi^+ \pi^-$,  with BRs determined by the mass of the $\gamma_d$ and of the virtual SM photon with which the $\gamma_d$ kinetically mixes. A $\gamma_d$ mass range from 0.1--2 GeV is considered. In the second scenario, a dark scalar $s_d$ decays to a pair of dark photons ($\gamma_d \gamma_d$) and each dark photon decays as described above. A  MC generator tool, called the LJ gun \cite{HiggsDisplacedLJ8TeVATLASpaper}, is used to generate these processes. For simplicity, the LJ gun samples are generated with only three dark-photon decay modes, $e^+ e^-$, $\mu^+ \mu^-$ or $\pi^+ \pi^-$, and the branching ratios are assigned based on the mass of the dark photons. The BRs of the $\gamma_d$ are determined with a single $\gamma_d$ sample. The $s_d$ masses considered range from $1$ GeV to $10$ GeV. The generated events are processed through the full ATLAS \textsc{Geant4}-based simulation chain. Additional $pp$ interactions in the same and nearby bunch crossings (pile-up) are included in the simulation.  All Monte Carlo samples are re-weighted to reproduce the observed distribution of the number of interactions per bunch crossing in the data. \\       

\begin{figure}[ht!]
\includegraphics[width=0.5\textwidth]{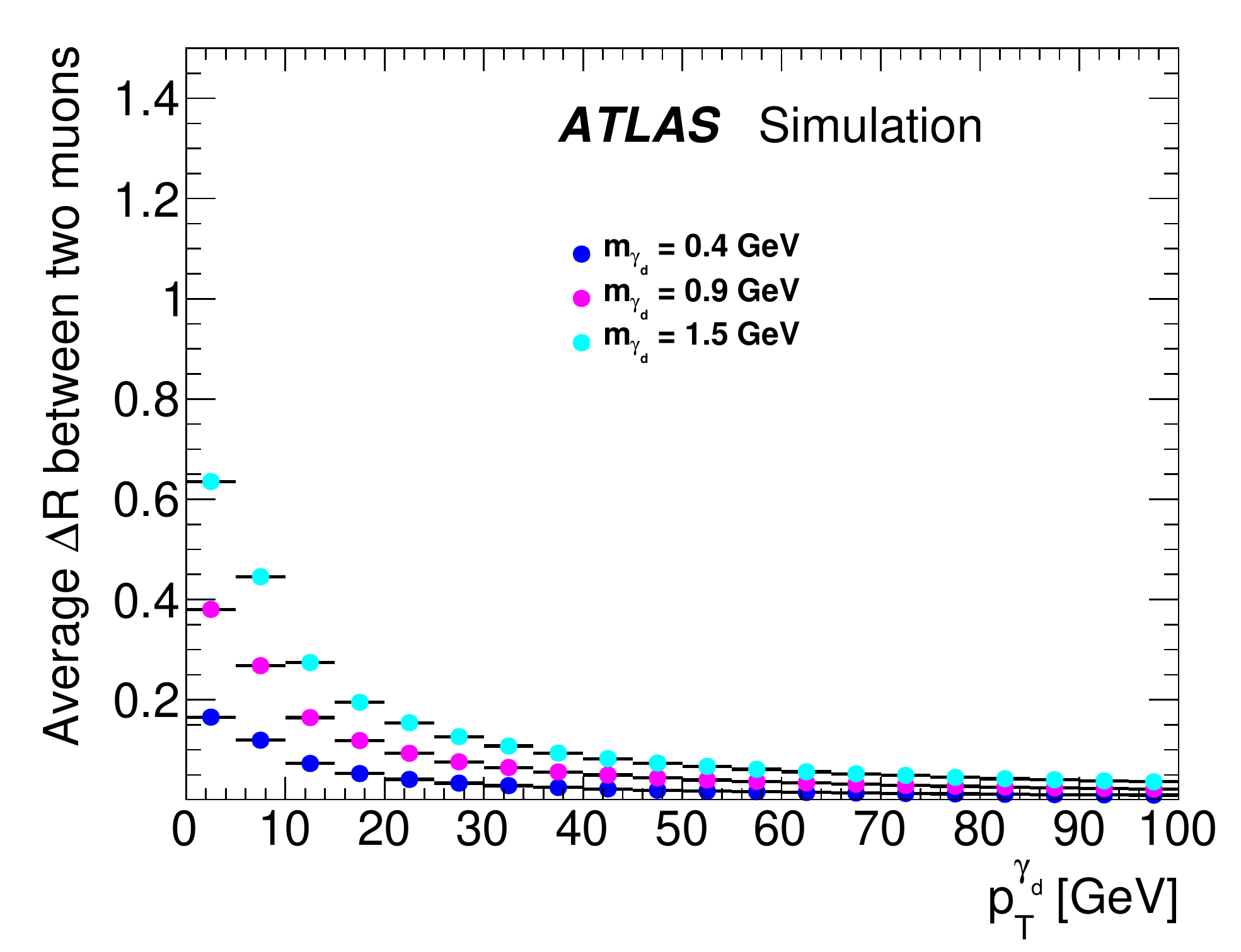}
\includegraphics[width=0.5\textwidth]{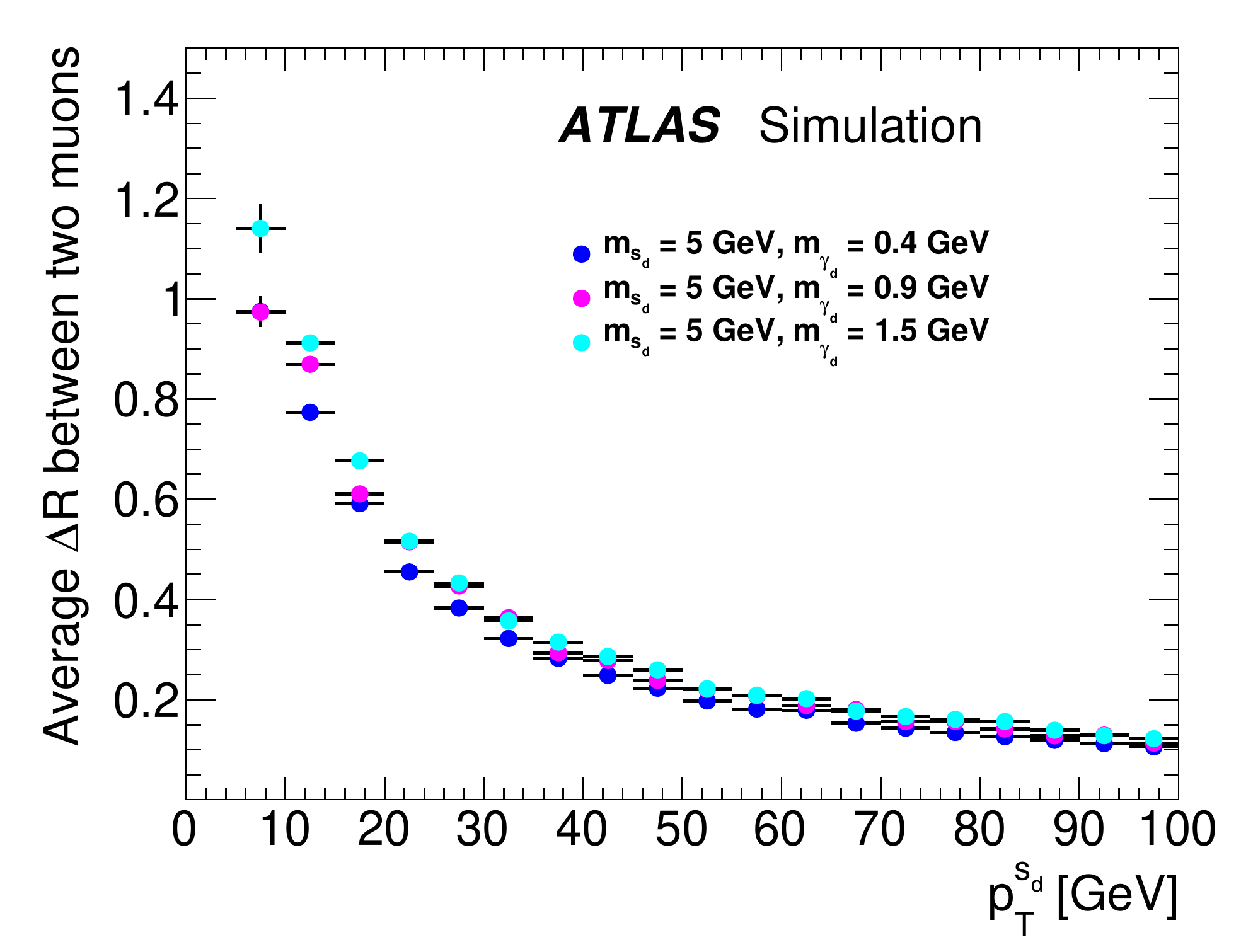}
\caption{The average separation between two truth muons in the LJ gun samples for various masses of the $\gamma_d$ (left) as a function of the $\pt$ of $\gamma_d$, and (right) with respect to the $\pt$ of a dark scalar particle $s_d$ with a mass of $5$ GeV.}
\label{fig:profdRVspt}
\end{figure}

\noindent The separation between the leptons depends on the mass and the boost of their parent $\gamma_d$. Figure \ref{fig:profdRVspt} shows the dependence of the average separation between muons, $\Delta R$, on the $\gamma_d$ or $s_d$ transverse momentum for various mass values.  In the left figure, the average $\Delta R$ is evaluated from the distribution for a given \pt slice of $\gamma_d$ decaying into two truth\footnote{The term "truth" is used to indicate objects derived directly from the Monte Carlo generator output, without considering the detector simulation.} muons. In the right figure, the average $\Delta R$ is evaluated from the distribution of six possible combinations of muon pairs from four muons for a given \pt bin of $s_d$ decaying into $\gamma_d$, where each $\gamma_d$ decays into a muon pair. The average $\Delta R$ decreases with increasing \pt of the dark particle. \\

\noindent Lepton-jet candidates are formed from ID tracks, energy clusters in the EM calorimeter and muons. In order to minimize the background from processes producing low-$\pt$ tracks, ID tracks are required to have a  minimum $\pt$ of 5 GeV. 
                       
%
\subsection{Lepton-jet reconstruction}
\label{sec:ljreco}
\noindent The reconstruction of lepton-jets starts by arranging the ID tracks from the primary vertex in order of decreasing $\pt$. The minimum $\pt$ of all tracks in the list is 5 GeV. Starting from the first track in the list, the next track in the list within $\Delta R = 0.5$ of the seed track is found. The four-momenta of the two tracks are summed to give the four-momentum of the lepton-jet candidate. Subsequent tracks within $\Delta R = 0.5$ of the lepton-jet candidate are added iteratively, recomputing the momentum sum at each step. This procedure is repeated until the track list is exhausted. Tracks that are added to a lepton-jet candidate are removed from consideration for subsequent  LJ candidates. Additional lepton-jets are built from the remaining tracks in the list following the same procedure. Each lepton-jet candidate contains at least two tracks.\\


\noindent The lepton-jet candidates are categorized as follows:

\begin{itemize}
\item{{\bf{Electron-jet (eLJ):}} If at least one reconstructed electron with \et > 10 GeV is found within $\Delta R = 0.5$  of the lepton-jet but no muons, the lepton-jet candidate is called an electron-jet (eLJ). Due to the spatial resolution of the EM calorimeter, the two electrons from an $\mathcal{O}(1~\text{GeV})$ $\gamma_d$ usually merge to form a single cluster. The two leading tracks must have \pt > 10 GeV, and all other tracks have \pt > 5 GeV.}     
          
\item{{\bf{Muon-jet (muLJ):}}  If at least two muons with \pt > 10 GeV are found within $\Delta R = 0.5$ of the lepton-jet but no electrons, the lepton-jet candidate is called a muon-jet (muLJ). The two muon tracks must have \pt > 10 GeV, and all other tracks have \pt > 5 GeV.}

\item{{\bf{Mixed-jet (emuLJ):}} If at least one reconstructed electron with \et > 10 GeV and at least one muon with \pt > 10 GeV is found within $\Delta R = 0.5$ of the lepton-jet cone, the lepton-jet candidate is called a mixed-jet (emuLJ). Mixed-jets are reconstructed from the $s_d$ producing two $\gamma_d$ pairs where one $\gamma_d$ decays to $e^+ e^-$ and the other to $\mu^+ \mu^-$. The leading track must have \pt > 10 GeV, and the sub-leading track and all other tracks have \pt >5 GeV.}  

\end{itemize}

\noindent Six categories of events are defined: those with two electron-jets (eLJ--eLJ), those with two muon-jets (muLJ--muLJ), and those with a mixed combination of jets (eLJ--muLJ, eLJ-- emuLJ,  muLJ--emuLJ, emuLJ--emuLJ).\\

\subsection{Lepton-jet reconstruction efficiency}
\label{sec:LJeffi}

\begin{figure}[h]
\includegraphics[width=0.5\textwidth]{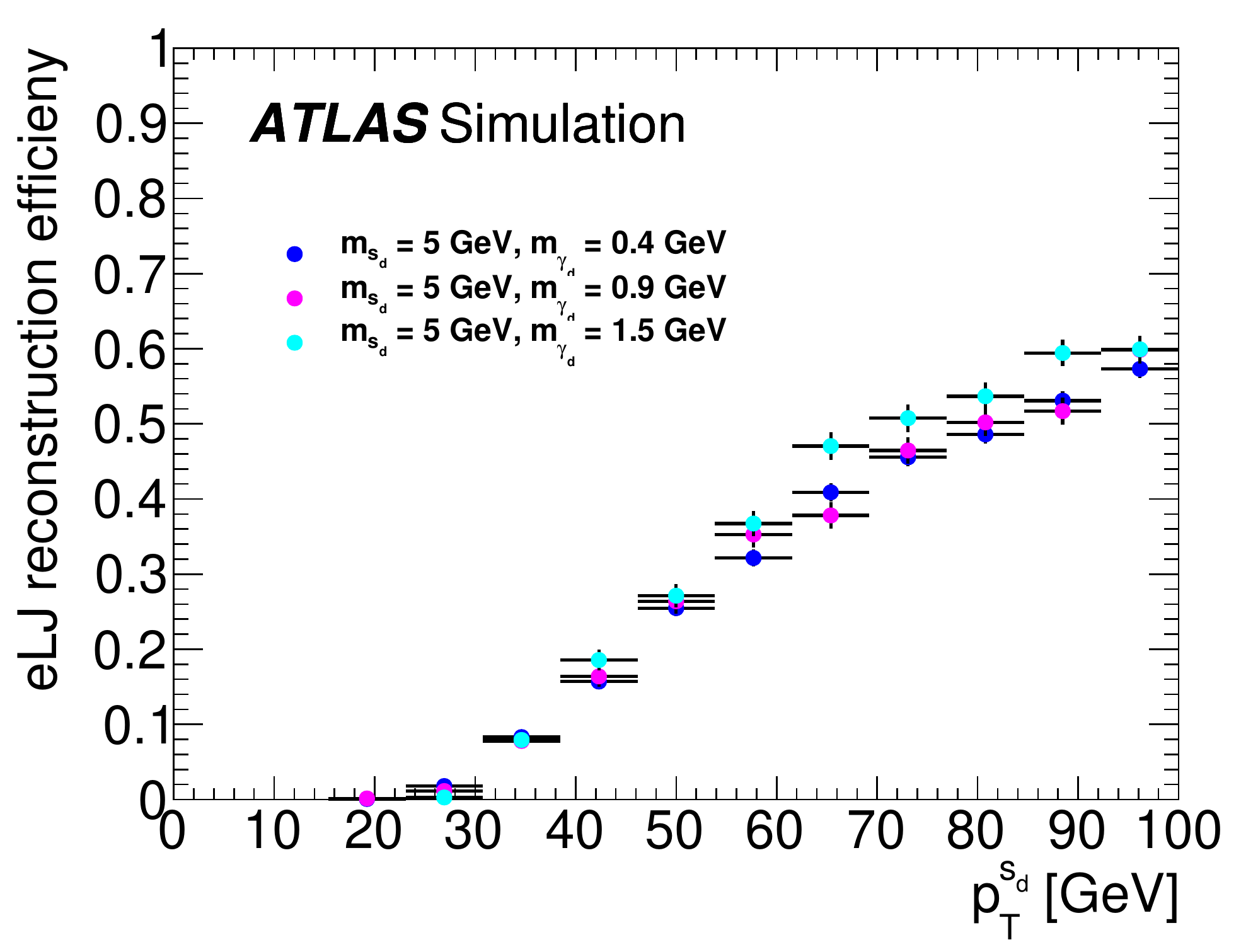}
\includegraphics[width=0.5\textwidth]{figures/fig_04b.pdf}
\caption{Efficiency of eLJ reconstruction as a function of $\pt$ of longitudinally (LP) or transversely (TP) polarized $\gamma_d$ for the process $\gamma_d \rightarrow e^+e^-$ (left) and for the process $s_d \rightarrow \gamma_d \gamma_d$, where at least one $\gamma_d$ decays to  $e^+ e^-$ (right).}
\label{fig:ELJrecoeffSinglegd}
\end{figure}

\noindent The characteristics of the reconstructed lepton-jets are studied using the LJ gun samples. The efficiency for lepton-jet reconstruction is $\pt$- and $\eta$- dependent. All efficiencies shown in this section are with respect to single lepton-jet events. The eLJ reconstruction efficiency (Figure \ref{fig:ELJrecoeffSinglegd}) is defined as the fraction of events having at least one truth $\gamma_d$ decaying to $e^{+}e^{-}$ which contain a reconstructed eLJ matched to the direction of the $\gamma_d$ ($s_d$) for single (double) $\gamma_d$ samples. The matching criterion is that at least one of the LJ’s clusters in the EM calorimeter lies within $\Delta R = 0.1$ of the truth  $\gamma_d$ ($s_d$) momentum direction. \\

\begin{figure}[h]
\includegraphics[width=0.5\textwidth]{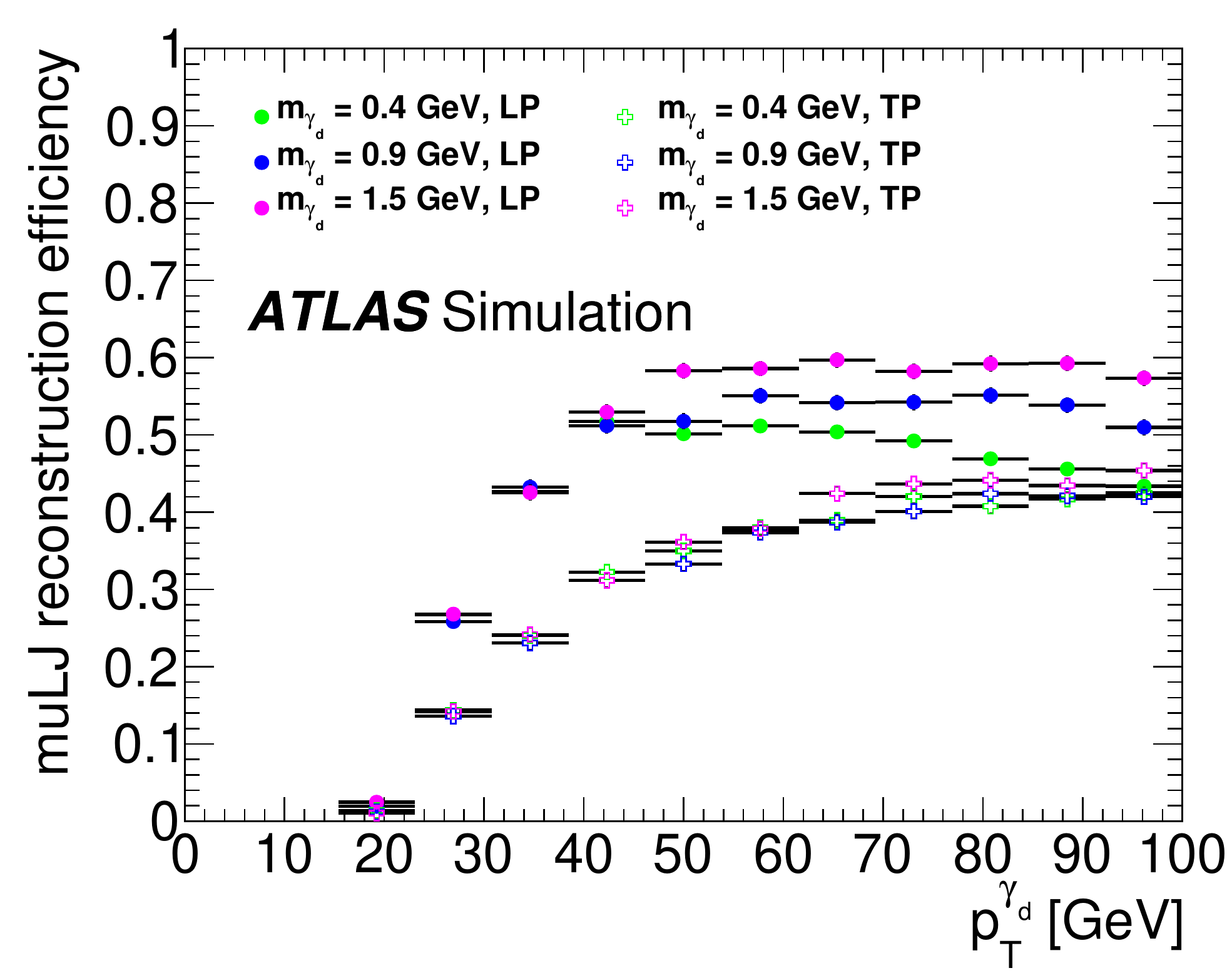}
\includegraphics[width=0.5\textwidth]{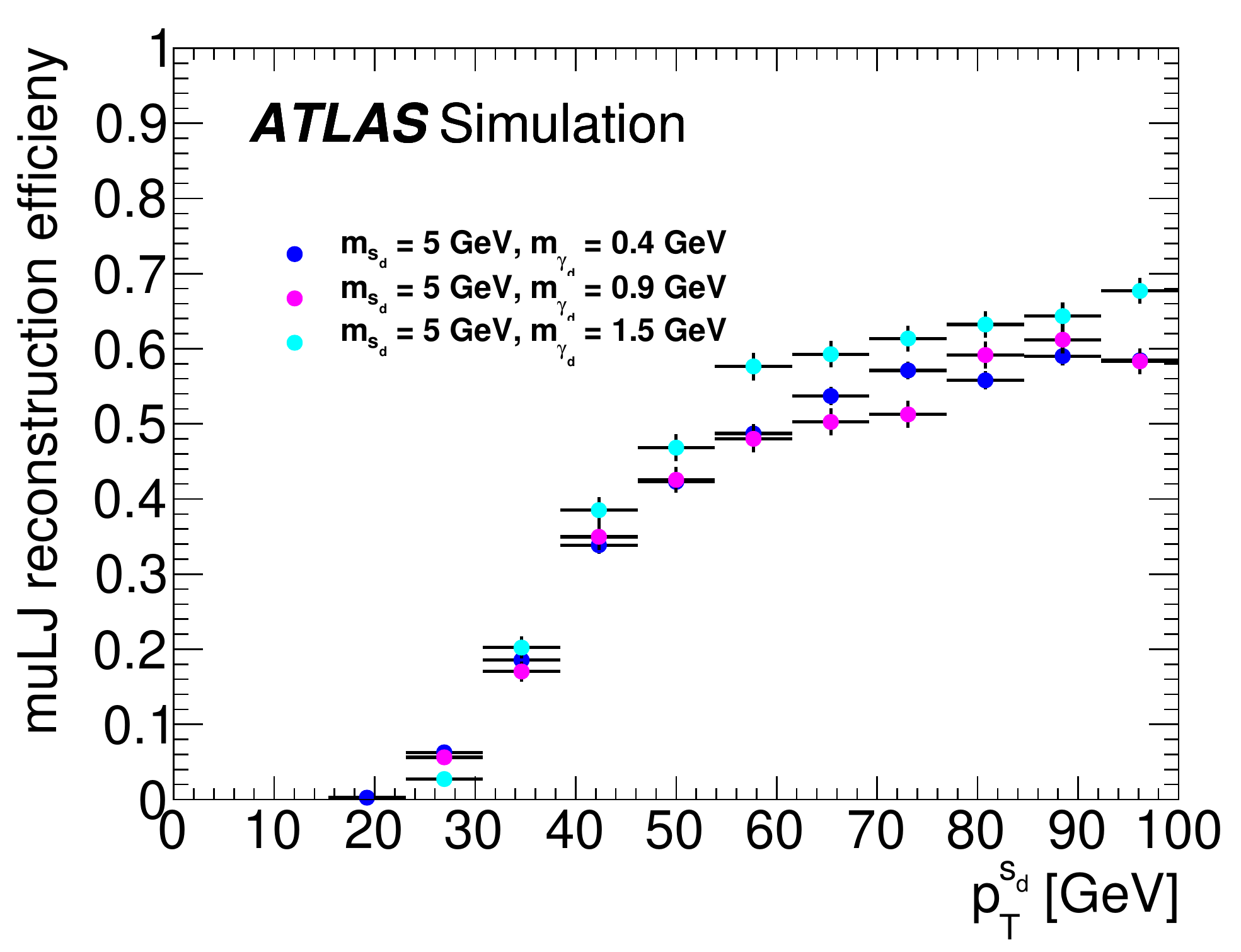}
\caption{Efficiency of muLJ reconstruction as a function of $\pt$ of longitudinally (LP) or transversely (TP) polarized $\gamma_d$ for the process $\gamma_d \rightarrow \mu^+ \mu^-$  (left) and for the process $s_d \rightarrow \gamma_d \gamma_d$, where at least one $\gamma_d$ decays to  $\mu^+ \mu^-$ (right).}
\label{fig:MuLJrecoeffSinglegd}
\end{figure}

\noindent The muLJ reconstruction efficiency (Figure \ref{fig:MuLJrecoeffSinglegd}) is defined as the fraction of events having at least one truth $\gamma_d$ decaying to $\mu^{+}\mu^{-}$ which contain a reconstructed muLJ matched to the direction of $\gamma_d$ ($s_d$) for single (double) $\gamma_d$ samples.  The matching criterion is that at least one of the \LJ's muons lies within $\Delta R = 0.1$ of the truth  $\gamma_d$ ($s_d$) momentum. \\

\noindent For both single and double $\gamma_d$ production, electron-jets have a higher reconstruction efficiency for higher $\pt$ dark photons. The rise in efficiency with the dark photon $\pt$ is due to the requiring at least two tracks with $\pt \ge 10$ GeV and at least one cluster with $\et \ge 10$ GeV. Longitudinally polarized (LP) dark photons \cite{HiggsDisplacedLJ8TeVATLASpaper} have a higher probability for the decay products to have unbalanced momenta than the transversely polarized (TP) ones. For eLJ, LP dark photons are more likely than TP ones to satisfy the  \pt > 10 GeV requirement as shown in Figure \ref{fig:ELJrecoeffSinglegd}. The slight decrease in muLJ efficiency at high $\pt$ in Figure \ref{fig:MuLJrecoeffSinglegd} is due to the smaller $\Delta R$ of  $\gamma_d$ decays along the  $\gamma_d$ momentum direction. For higher $p_T$ $\gamma_d$, the LP muLJ decay products are more often reconstructed as a single muon in the MS. As shown in  Figure \ref{fig:MuLJrecoeffSinglegd} (right), for the double $\gamma_d$ case, the muLJ reconstruction efficiency improves with the dark-photon $\pt$, as only one of the $\gamma_d$ needs to be reconstructed.

\subsection{Background rejection at the lepton-jet level}
\label{sec:LJbkg}

\noindent The reconstructed sample of lepton-jets includes SM backgrounds, mostly hadronic jets that are misidentified as lepton-jets. The variables that discriminate between signal and background processes are based on the characteristics of the reconstructed lepton-jets. Electron LJs from dark particles are expected to have a different isolation (defined below) around the electron tracks in the ID. They are also expected to have different shower shapes in the EM calorimeter when compared to hadronic jets from SM processes, due to the presence of multiple collimated electrons. The muons in the muLJs are more isolated in the ID and the calorimeter than the muLJ backgrounds from multijet processes. In Sections \ref{sec:eljvariables} to \ref{sec:emuljvariables}, all the variables considered for eLJ, muLJ and emuLJ are listed. Section \ref{sec:finalsel} explains how the cut values are optimized. The shape of the distribution of each variable is qualitatively compared between data and simulation in a $Z$+jets sample. A few of these variables are shown in Figures  \ref{fig:track} and \ref{fig:emfrac}. The $Z$+jets events are selected by requiring two opposite sign leptons, where the invariant mass of two leptons is within 10 GeV of the $Z$ boson mass window.

\subsubsection{eLJ variables}
\label{sec:eljvariables}

\begin{itemize}
\item{{\bf{Track isolation:}} Track isolation (Figure \ref{fig:track}, left ) is defined as the ratio of the scalar sum of the $\pt$ of the ID tracks within $\Delta R = 0.5$ around the eLJ direction, excluding the EM cluster-matched tracks, to the eLJ $\pt$. Each ID track used for the isolation calculation must have $\pt \ge 1$ GeV and $|\eta| \le 2.5$. To reduce pile-up dependence, each ID track must pass transverse and longitudinal impact parameter requirements, $|d_0| < 1$ mm and $|z_0 \sin \theta| < 1.5$ mm, respectively. The tracks matched to a cluster in the EM calorimeter are defined to be the ID tracks with $\pt \ge 5$ GeV that either lie within $\Delta R = 0.05$ from the cluster or are among the two tracks closest to the cluster.}

\begin{figure}[ht!]
\includegraphics[width=0.5\textwidth]{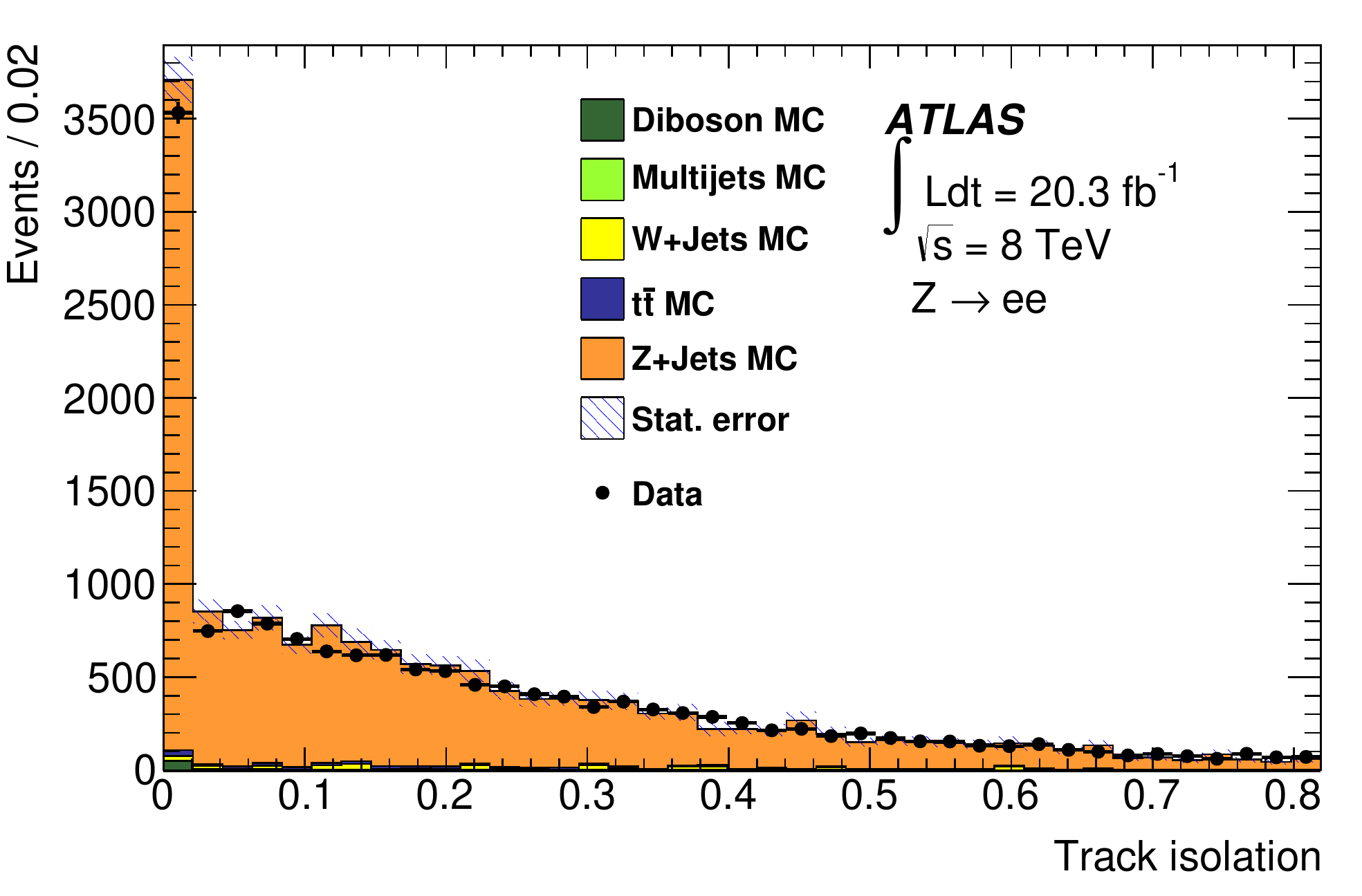}
\includegraphics[width=0.5\textwidth]{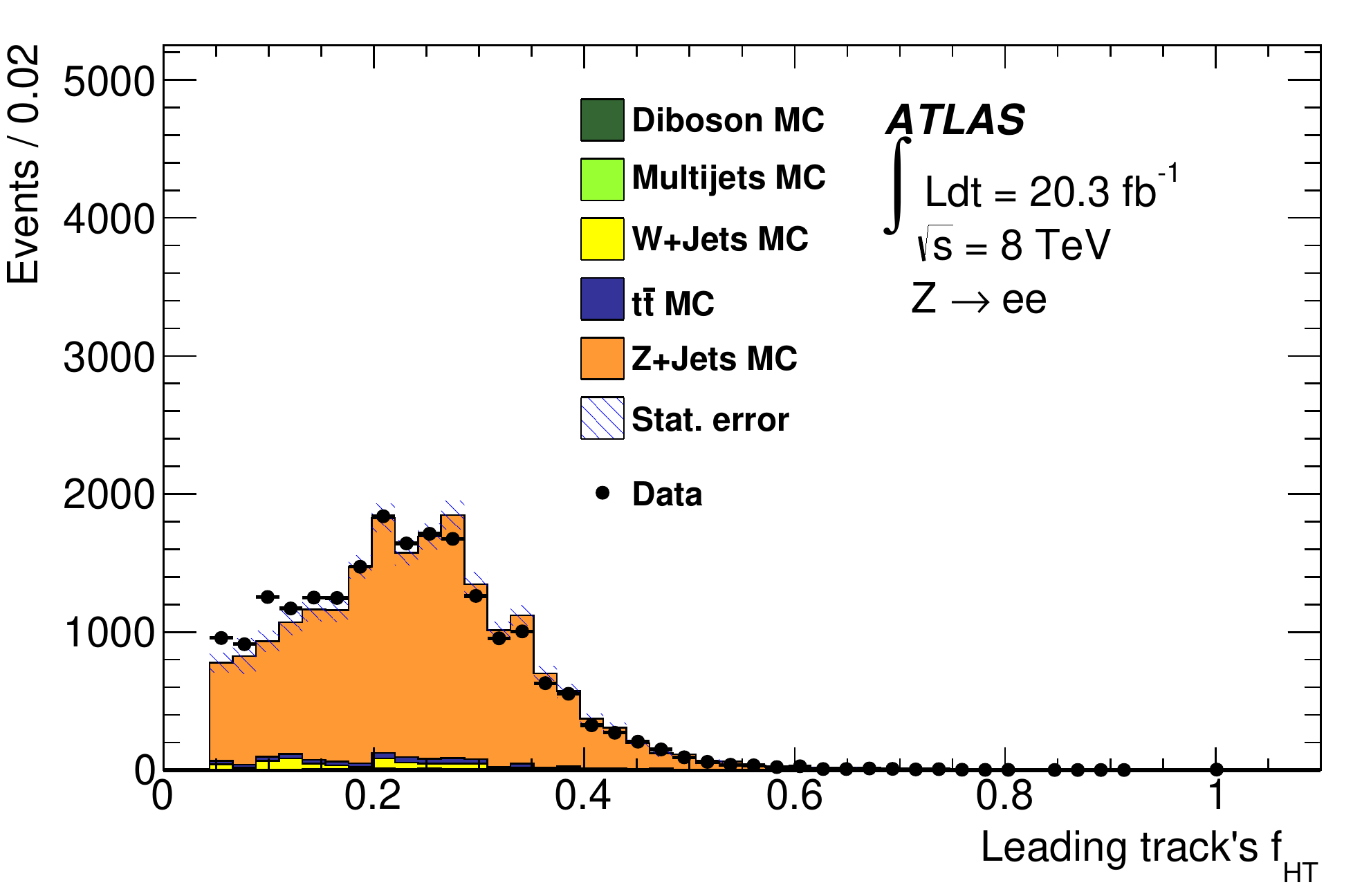}
\caption{Distributions of track isolation and the fraction of high-threshold TRT hits $f_{\text{HT}}$ of the leading track in a $Z \to e^+ e^-$ sample.}
\label{fig:track}
\end{figure} 

\item{{\bf{Fraction of high-threshold TRT hits, $f_{\text{HT}}$ (Figure \ref{fig:track}, right):}} Electrons deposit more energy in the TRT than pions due to transition radiation. The ratio of the number of high-threshold TRT hits (6 keV) on the track to the total number of TRT hits is a robust discriminating variable to identify electrons in an eLJ.}

\item{{\bf{Energy of the strip with maximal energy deposit, $E_{s1}^{\text{max}}$ (Figure \ref{fig:emfrac}, left):}} The first sampling layer of the EM calorimeter has a finer granularity in $\eta$ compared to the second sampling layer. This granularity is used to distinguish between electron and $\pi^{0}$ showers. An electron tends to deposit its energy in a single strip, while the $\pi^0 \rightarrow \gamma\gamma$ decays tend to share energy between two strips, providing a discriminant for $\pi^0$ rejection.} 

\item{{\bf{Fraction of energy deposited in the third sampling layer of the EM calorimeter, $f_{s3}$:}} Electrons deposit most of their energy into the second sampling layer of the EM calorimeter, leaving only a small amount of energy in the third sampling layer. Hadrons deposit most of their energy in the hadronic section of the calorimeters and a small deposition in the second sampling layer of the electromagnetic calorimeter. Furthermore, hadrons deposit a relatively larger amount of energy in the third sampling layer compared to electrons. The fraction of energy found in the third sampling layer of the electromagnetic calorimeter is used to discriminate between electrons and hadrons.}

\item{{\bf{Electromagnetic energy fraction, $f_{\text{EM}}$ (Figure \ref{fig:emfrac}, right):}} This is the fraction of the cluster's total transverse energy found in the EM calorimeter. An eLJ is expected to have a larger EM fraction than hadronic jets from SM processes. $f_{\text{EM}}$ can be negative because the calibrated energy deposition in the hadronic calorimeter can be negative due to noise subtraction.}
\end{itemize}

\subsubsection{muLJ variables}
\label{sec:muljvariables}

\begin{itemize}
\item{{\bf{Track isolation:}} The ratio of the scalar sum of the $\pt$ of the ID tracks, excluding the muon tracks, within $\Delta R = 0.5$ around the muLJ direction, to the $\pt$ of the muLJ. Each ID track used for the isolation calculation must have $\pt \ge 1$ GeV and $|\eta| \le 2.5$. To reduce pile-up dependence, each ID track must pass transverse and longitudinal impact parameter requirements,  $|d_0| < 1$ mm and $|z_0 \sin \theta| < 1.5$ mm, respectively. The muon tracks are defined as the ID tracks with a tighter $\pt$ requirement, $\pt \ge 5$ GeV, which are matched to the fitted muon tracks within $\Delta R = 0.05$.} 
\item{{\bf{Calorimeter isolation:}} The ratio of the total transverse energy in the calorimeter within $\Delta R = 0.2$  from the leading muon of a muLJ to the  $\pt$ of that muon.}

\end{itemize}

\subsubsection{emuLJ variables}
\label{sec:emuljvariables}

\begin{itemize}
\item{{\bf{Track isolation:}}  The ratio of the scalar sum of the $\pt$ of the ID tracks (excluding the electron tracks and muon tracks within $\Delta R = 0.5$ around the emuLJ direction) to the emuLJ $\pt$.}

\item{{\bf{Energy of the strip with maximal energy deposit, $E_{s1}^{\text{max}}$:}} As described in Section~\ref{sec:eljvariables}.}

\item{{\bf{Fraction of energy deposited in third sampling layer of EM calorimeter, $f_{s3}$:}} As described in Section~\ref{sec:eljvariables}.}

\item{{\bf{Hadronic leakage, $E_{\text{T}}^{\text{had}}$:}} The transverse energy of the electron deposited in the first sampling layer of the hadronic calorimeter. The emuLJ is expected to have a small hadronic contribution due to the presence of a muon within the cone. The $E_{\text{T}}^{\text{had}}$ is more sensitive than the $f_{\text{EM}}$ variable for the emuLJ case.}
\end{itemize}

\begin{figure}[ht!]
\includegraphics[width=0.5\textwidth]{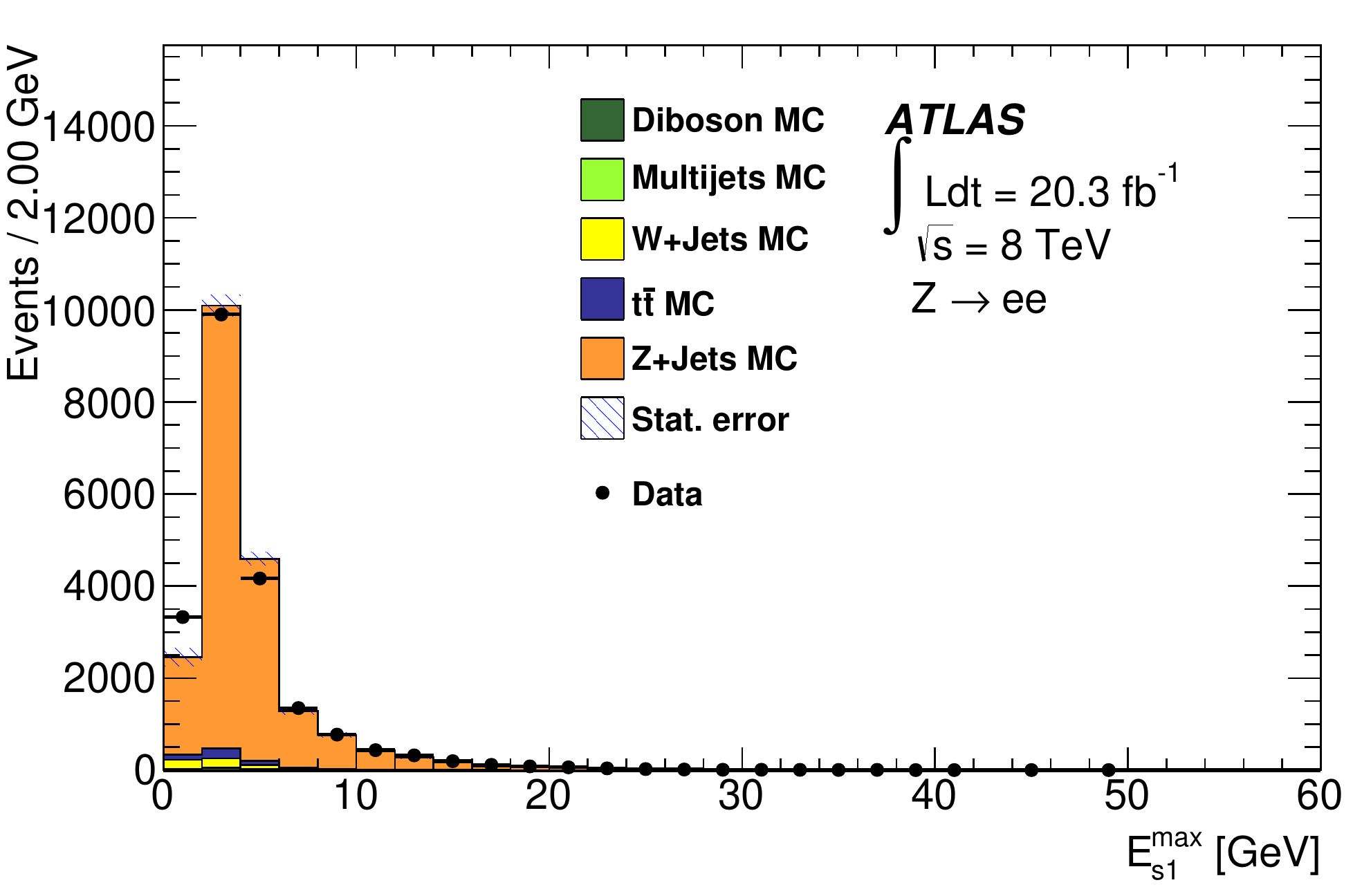}
\includegraphics[width=0.5\textwidth]{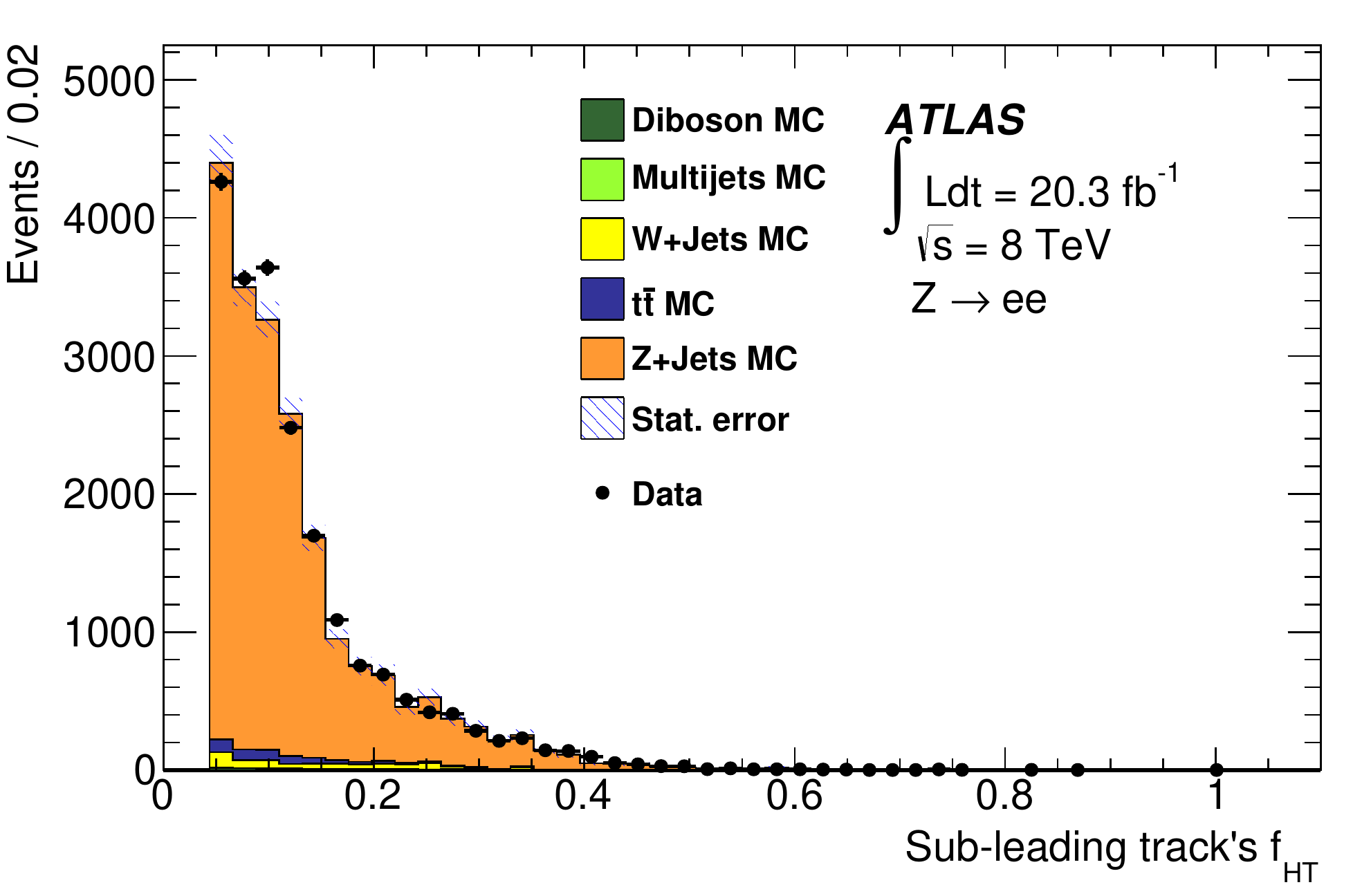}
\caption{Distributions of the energy of the strip with maximal energy deposit $E_{s1}^{\text{max}}$ and the electromagnetic energy fraction  $f_{\text{EM}}$ in a $Z \to e^+ e^-$ sample.}
\label{fig:emfrac}
\end{figure}

\subsubsection{LJ variables optimization}
\label{sec:finalsel}

Cuts on the variables described in Sections \ref{sec:eljvariables} to \ref{sec:emuljvariables} are used to suppress SM backgrounds. A multi-dimensional space of cuts is explored. The optimal cut values are selected by maximizing the significance expression~\cite{CutOptimization},
\begin{eqnarray}
\text{Significance} = \frac{n_{\text{S}}}{1 +  \sqrt{n_{\text{B}}}} ,
\label{eqn:signif}
\end{eqnarray}
where $n_{\text{S}}$ denotes the expected number of signal events, $n_{\text{S}} = \epsilon_{\text{s}} \times \sigma \times \mathcal{L}$, and $n_{\text{B}}$ is the number of background events passing the selection cuts. The variable $\epsilon_{\text{s}}$ represents the signal efficiency formed by the product of the LJ reconstruction efficiency (Section \ref{sec:LJeffi}), the efficiency of requirements on the discriminating variables and the LJ trigger efficiency. The optimized cuts for all LJ variables are given in Table~\ref{tab:cutvalues}. The optimization is performed for the efficiency of two-LJ events, which is obtained from the product of single LJ efficiencies. The cut efficiency variation as a function of the dark-photon mass is studied for the eLJ--eLJ, muLJ--muLJ and eLJ--muLJ channels in case of $2 \gamma_d$ production. The variation is found to be small. For $2 s_d$ production the cut efficiency variation with the dark-photon mass can be large when one of the $\gamma_d$ originating from $s_d$ decays into hadrons and the other into a lepton-pair. The size of the variation in this case depends on the branching ratio for $\gamma_d$ decay to hadrons. Table \ref{tab:cutvalues2} summarizes the signal acceptance $\times$ efficiencies for the $m_{\gamma_d} = 0.4$ GeV benchmark, where efficiencies for events passing all the cuts are shown. These efficiencies take into account the trigger selection, primary vertex selection, lepton-jet reconstruction, and efficiencies of discriminating variable cuts, and do not include the BRs for $\gamma_d$ decays into $e^+ e^-$ or $\mu^{+} \mu^{-}$ pairs in each channel. The difference between efficiencies for the Higgs-portal and SUSY-portal models are driven by the boost of the dark photon which depends on the mass of the parent particles, the squark or Higgs boson.

\begin{table}[ht!]
\centering
\caption{Finalized set of cut values on the discriminating variables of eLJ, muLJ and emuLJ.}
\begin{tabular}{| c | c | c |}
\hline 
eLJ & muLJ & emuLJ \\
\hline
$E_{s1}^{\text{max}}$ \textgreater~ 0.5 GeV	&  								&	$E_{s1}^{\text{max}}$ \textgreater~ 3 GeV \\
track isolation \textless~ 0.04 	& track isolation \textless~ 0.25 		&	track isolation \textless~ 0.1 \\
$f_{\text{HT}}$ \textgreater~ 0.14 		& calorimeter isolation \textless~ 0.15 		&	$E_{\text{T}}^{\text{had}}$ \textless~ 1 GeV \\
$f_{\text{EM}}$ \textgreater~ 0.99		&  								&	$f_{s3}$ \textless~ 0.015	\\
$f_{s3}$ \textless~ 0.015		&  								&	\\
\hline
\end{tabular}
\label{tab:cutvalues}
\end{table}

\begin{table}[ht!]
\footnotesize
\centering
\caption{Acceptance $\times$ efficiency corresponding to $m_{\gamma_d}$ = 0.4 GeV for all six channels in the SUSY-portal and Higgs-portal topologies.}
\begin{tabular}{| c | c | c | c | c | c | c |}
\hline
State & eLJ--eLJ & muLJ--muLJ  & eLJ--muLJ & eLJ--emuLJ & muLJ--emuLJ & emuLJ--emuLJ\\
\hline
\underline {SUSY-portal} & & & & & & \\
2 \gammad + X & 4.4 $\pm$ 0.2 \% &    6.4 $\pm$ 0.3\% & 3.4 $\pm$ 0.2 \% & - & - & - \\
2 $(s_d \rightarrow \gamma_d \gamma_d)$ + X & 6.3 $\pm$ 0.4 \% & 25.1 $\pm$ 0.7\% & 7.2 $\pm$ 0.3 \% & 4.0 $\pm$ 0.2 \% & 8.1 $\pm$ 0.3 \% & 7.1 $\pm$ 0.3 \% \\
\underline {Higgs-portal} & & & & & & \\
2 \gammad + X & 0.23 $\pm$ 0.02 \% & 1.31 $\pm$ 0.04 \% & 0.20 $\pm$ 0.01 \% & - & - & - \\
2 $(s_d \rightarrow \gamma_d \gamma_d)$ + X & 0.03 $\pm$ 0.02 \% & 0.50 $\pm$ 0.07 \% & 0.08 $\pm$ 0.01 \% & 0.05 $\pm$ 0.01 \% & 0.22 $\pm$ 0.03 \% & 0.08 $\pm$ 0.02 \% \\
\hline
\end{tabular}
\label{tab:cutvalues2}
\end{table}


\section{Background estimation at the event level}
\label{sec:bkg_est}

\noindent  All background contributions are estimated using the data-driven ABCD-likelihood method (Section \ref{sec:ABCD}), except for diboson top-quark pair ($t \bar{t}$), which are determined from MC simulations. The diboson estimation includes $\gamma^{*}$ production with any mass. The MC simulations for other SM processes are used only to investigate the shapes of the distributions of the LJ variables, and not for the background evaluation. \\

\noindent Various SM processes can mimic a LJ signal  due to hadrons being misidentified as leptons. The following MC samples are considered: hadronic multijet
events, $\gamma$ + jets events, $W (\rightarrow \ell \nu)$ + jets, $Z (\to \ell^+ \ell^-)$ + jets, $t\bar{t}$ and diboson
($WW$, $WZ$, $ZZ$, $\gamma\gamma$) events. \textsc{Pythia} 8 is used to generate these samples except for $t\bar{t}$, $WW$, $WZ$, $ZZ$ for which
MC@NLO~\cite{MCatNLO} is used. The contribution from $WZ$ and $ZZ$ backgrounds, when one of the bosons is off-shell, is modelled with \textsc{Sherpa}~\cite{Sherpa}. \\

Of the backgrounds considered, only the hadronic multijet, $\gamma$ + jets and $Z (\to \ell^+ \ell^-)$ + multijet events contribute significantly. The contribution from $t\bar{t}$ is negligible. The hadronization of the multijet, photon + jets and $W/Z$ + jets samples is done with \textsc{Pythia} 8 using the CT10 NLO \cite{CT10} PDF set. For the underlying event, the AUET2 set of tuned parameters \cite{mc11tune} is used. As with the signal MC samples, the SM MC samples included the effect of multiple $pp$ interactions per bunch crossing and are assigned an event weight such that the distribution of the number of $pp$ interactions matches that in data. All MC events are processed  with the \textsc{Geant4}-based ATLAS detector simulation~\cite{Geant1,atlas_simulation} and then analysed with the standard ATLAS reconstruction software.

\subsection{Low-mass Drell--Yan}

\noindent The contribution from low-mass Drell--Yan events $\gamma^* (\to \ell^+ \ell^-)$ + jets in the ranges $2 < m_{\ell\ell} < 8$ GeV and $10 < m_{\ell\ell} < 60$ GeV is investigated using MC simulation (\textsc{Sherpa}). This contribution is small because tracks from $\gamma^{*}$ have a soft \pt spectra, while the analysis requires that the tracks present in a lepton-jet to have \pt > 10 GeV. Furthermore, the requirement of two lepton-jets per event makes this background small. The remaining background is taken into account in the ABCD data-driven estimation (Section \ref{sec:ABCD}), as the ABCD plane is defined based on the sub-leading lepton-jet, which for $\gamma^{*}$ + jets  is predominantly a hadronic jet. The $\gamma^{*}$ + $\gamma^{*}$ background is evaluated from \textsc{Sherpa} MC simulations, and is subtracted from the event counts in the $A$, $B$, $C$ and $D$ regions while doing the fit for the hadronic jet background (see next section). \\

%
\subsection{Background estimation with the ABCD-likelihood method}
\label{sec:ABCD}

\noindent An ABCD-likelihood method is used to determine the lepton-jet backgrounds from SM processes. The method uses two nearly uncorrelated variables which have good discriminating power against background. By making a cut in each of the two variables,
four non-overlapping regions are defined, of which one is the signal region, labelled region $A$. Ideally, most signal events are concentrated in region $A$, while the other regions $B$, $C$ and $D$ are the control regions dominated by background events.
Using the event yields in the four regions, the background in the signal region is determined. Background processes contribute to the signal region selection because jets can be misidentified as lepton-jets.
The dominant jet background originates from multijet processes, while $W$/$Z$/$\gamma$/$\gamma^{*}$ + jets production accounts for less than 1\%  in the signal region. The two discriminating variables used in the ABCD data-driven background estimation show a small correlation of about 6\% in all LJ pairs except emuLJ--emuLJ.The emuLJ-emuLJ channel has fewer events, and the correlation between the variables is between 10\% and 38\%. All selections (Section \ref{sec:eventsel}) are applied, except for the two discriminating variables which are used for the ABCD-likelihood method. \\

\noindent  The ABCD-likelihood method estimates the expected background by fitting a likelihood function to the observed number of events in each of the four regions. The predicted event rates in each region are defined as follows:

\begin{itemize}
	\item $\mu_{A} = \mu^{U} + \mu + \mu_A^{K}$ 
	\item $\mu_{B} = \mu^{U}\tau_{B} + \mu b + \mu_B^{K}$
	\item $\mu_{C} = \mu^{U}\tau_{C} + \mu c + \mu_C^{K}$
	\item $\mu_{D} = \mu^{U}\tau_{B}\tau_{C} + \mu d + \mu_D^{K}$ 
\end{itemize}

\noindent Here, $\mu$ is the signal yield,  $\mu^{U}$ is the background yield from multijet and $W$/$Z$/$\gamma$/$\gamma^{*}$ + jets production in association with hadronic jets, while $b$, $c$, and $d$ describe the signal contamination (fraction of signal events in the control regions $B$, $C$ and $D$, divide by the fraction in the signal region). The variables $\tau_{B}$ and $\tau_{C}$ are the nuisance parameters that describe the ratio of the background expectation in the control region to the background expectation in the signal region. Lastly $\mu_A^{K}, \mu_B^{K}, \mu_C^{K}$ and $\mu_D^{K}$ represent the sum of the diboson and $\ttbar$ backgrounds, which are estimated from simulation since they are very small. These are taken as fixed parameters while doing the likelihood fit. The signal and background yields, as well as the values of the nuisance parameters, are obtained from the maximum-likelihood fit to the observed number of events ($n_A$, $n_B$, $n_C$, and $n_D$) in the four regions. The overall likelihood function is the product of the four likelihood functions in the four regions:

\begin{equation}
L(n_A, n_B, n_C, n_D | \mu, \mu_{U}, \tau_{B}, \tau_{C}) = \prod_{i=A,B,C,D} \frac{e^{-\mu_i} \mu_i^{n_i}}{n_i !}.
  \label{eqn:abcd-lhood}
\end{equation}

\begin{figure}[ht!]
	\subfloat[][]{
		\includegraphics[width=0.5\textwidth]{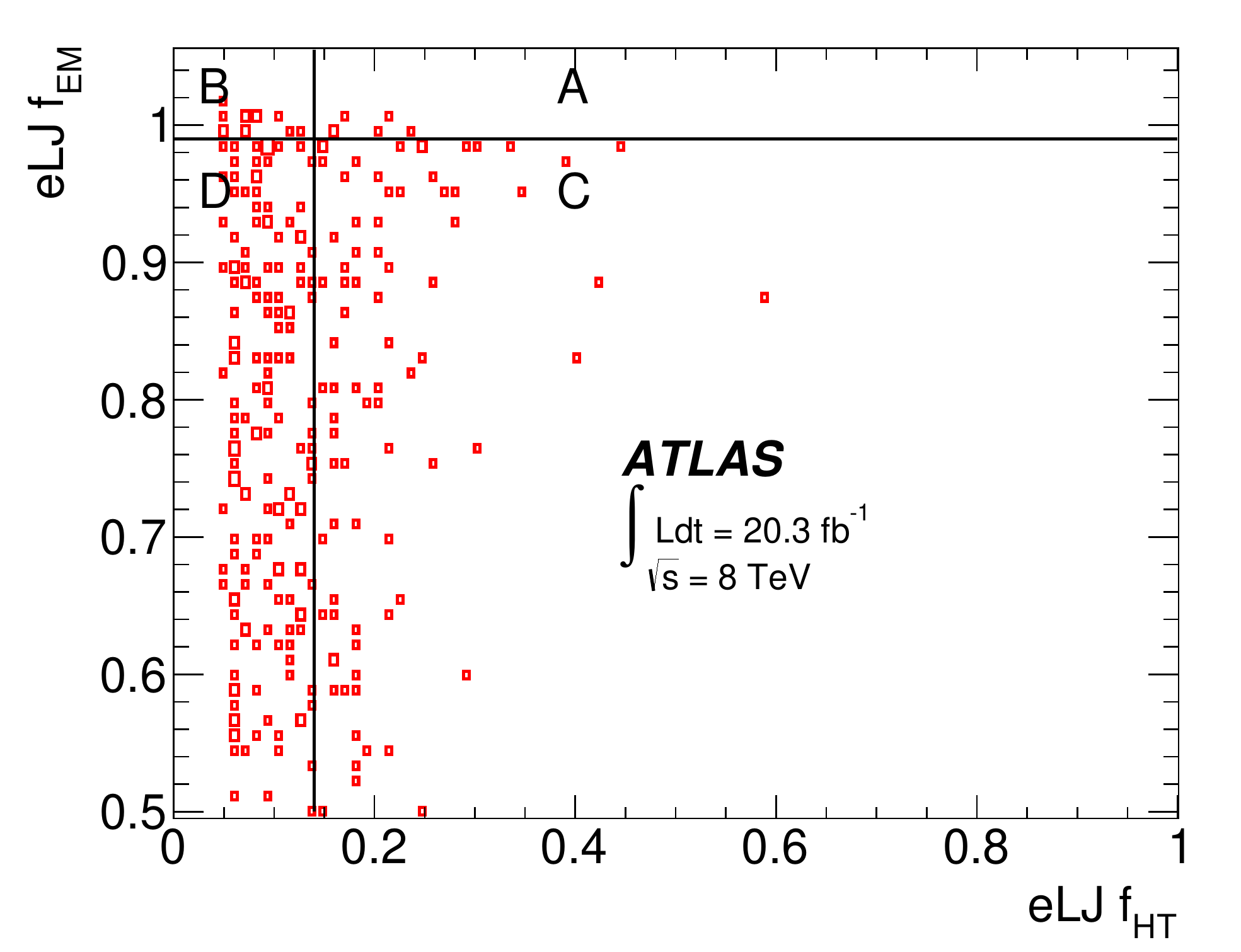}     
		\label{fig:ElEl_correlationSigStream}
	}
	\hfill
	\subfloat[][]{
		\includegraphics[width=0.5\textwidth]{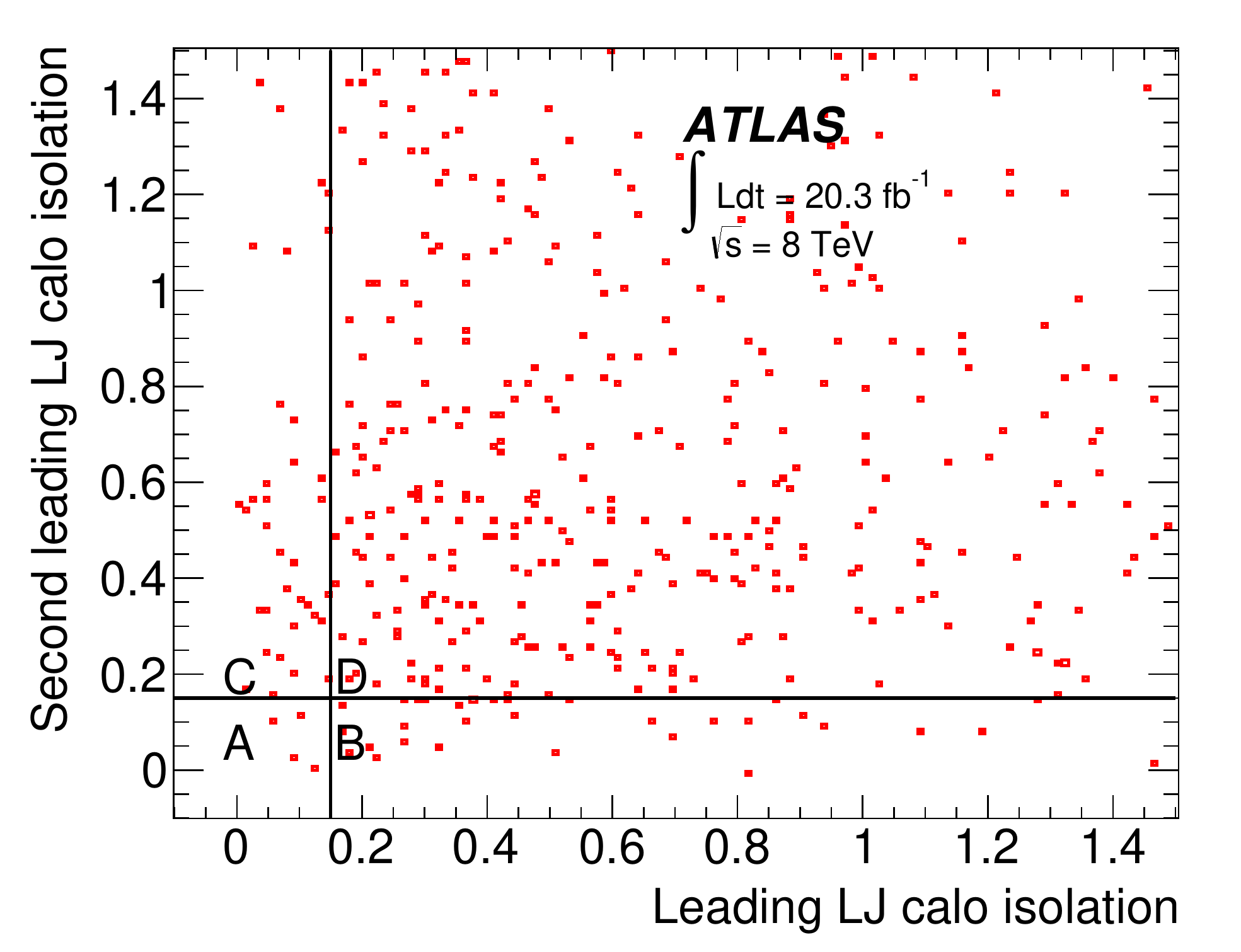}
		\label{fig:MuMu_correlationSigStream}
	} \\
	\subfloat[][]{
		\includegraphics[width=0.5\textwidth]{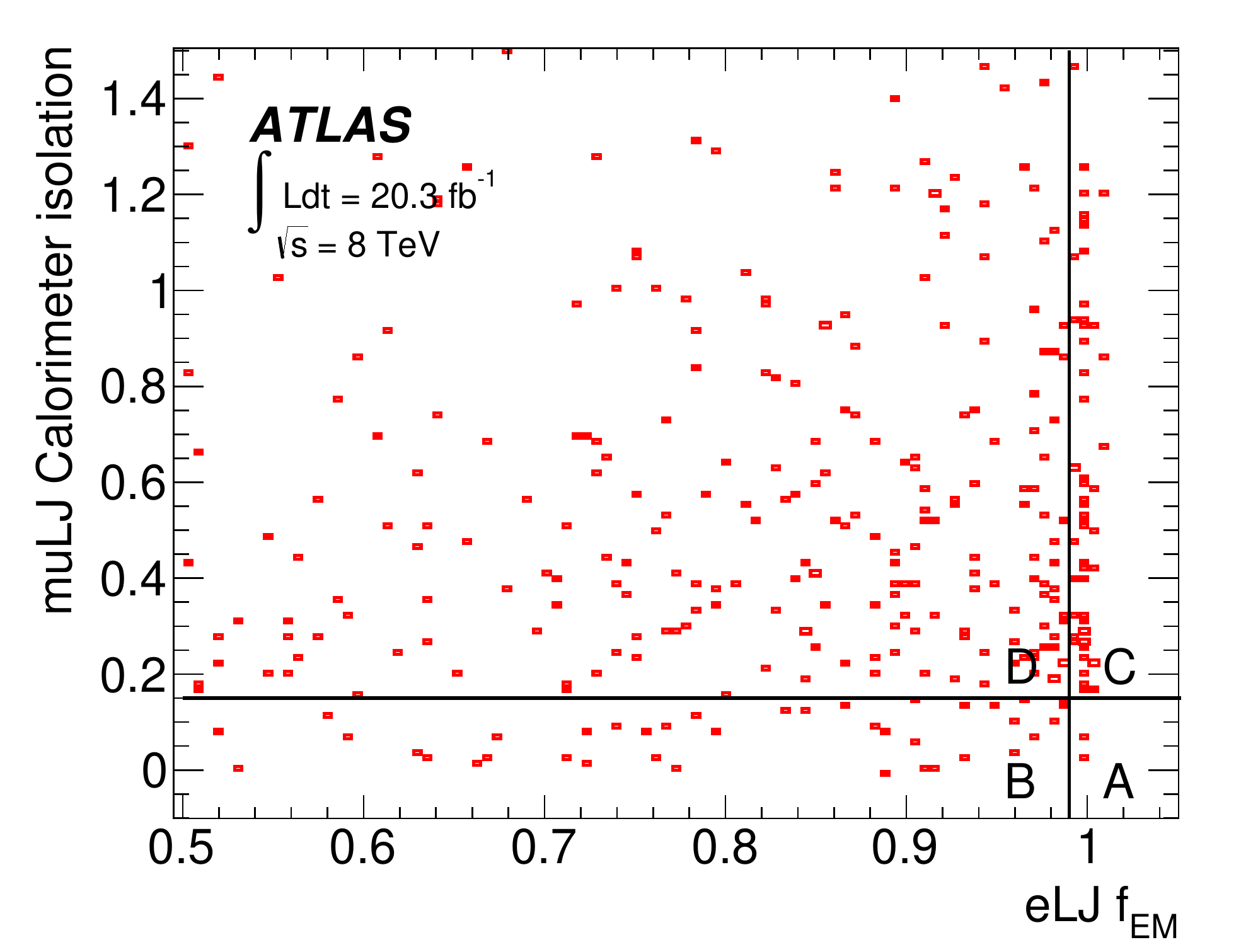}
		\label{fig:ElMu_correlationSigStream}
	}
	\hfill
	\subfloat[][]{
		\includegraphics[width=0.5\textwidth]{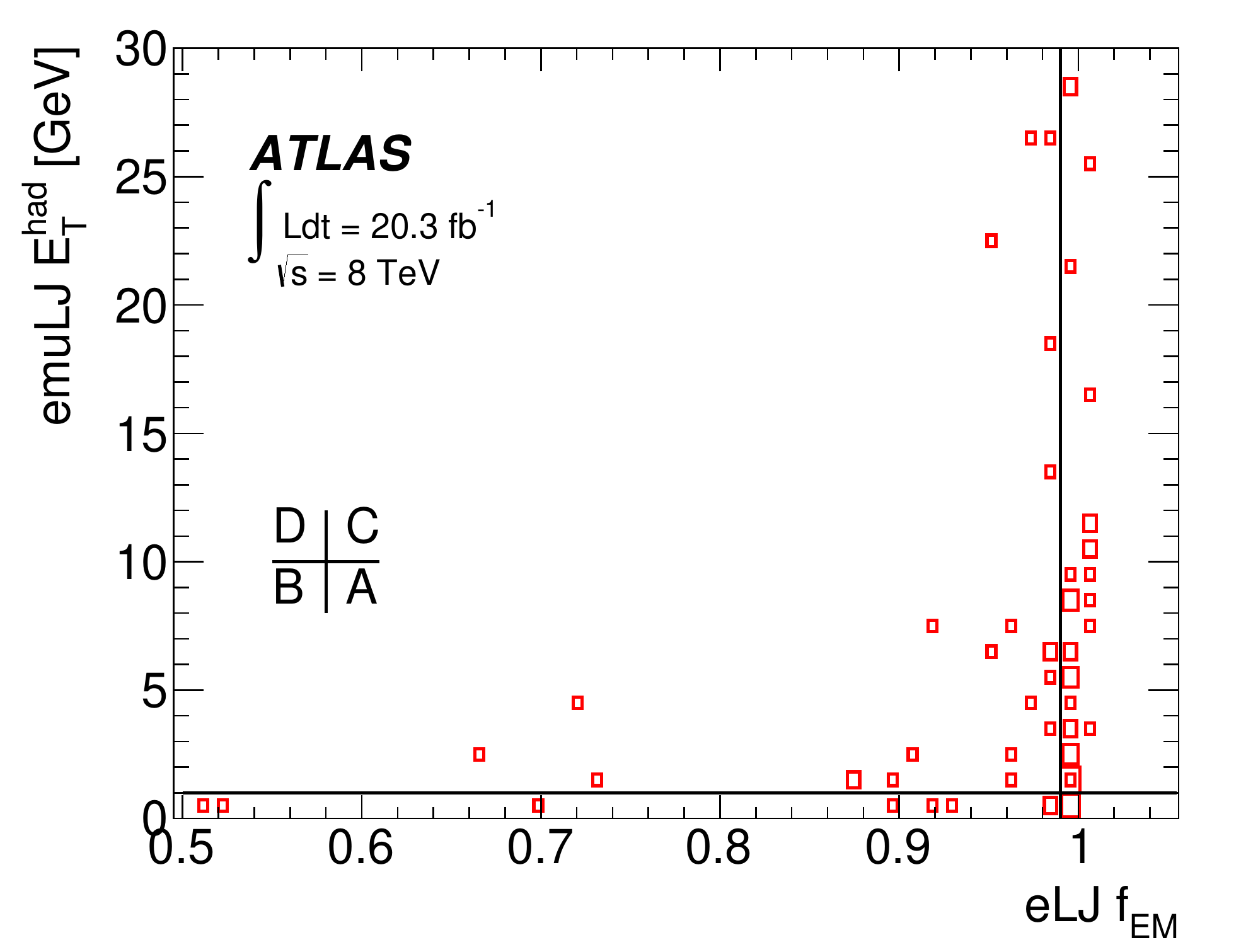}        
		\label{fig:ElEMu_correlationSigStream}
         }
         \hfill
          \subfloat[][]{
		\includegraphics[width=0.5\textwidth]{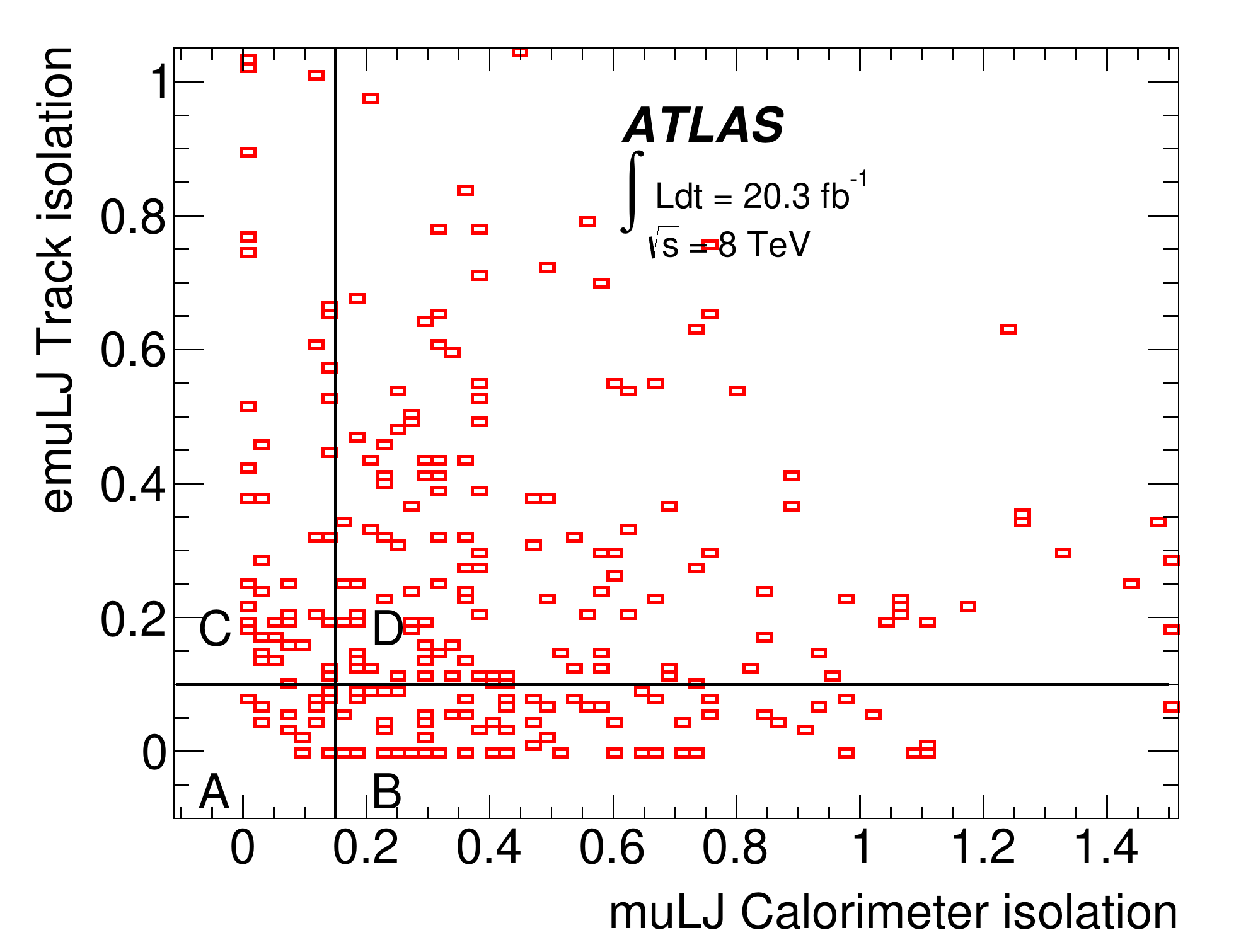}     
		\label{fig:MuEMu_correlationSigStream}
	}
	\hfill
	\subfloat[][]{
		\includegraphics[width=0.5\textwidth]{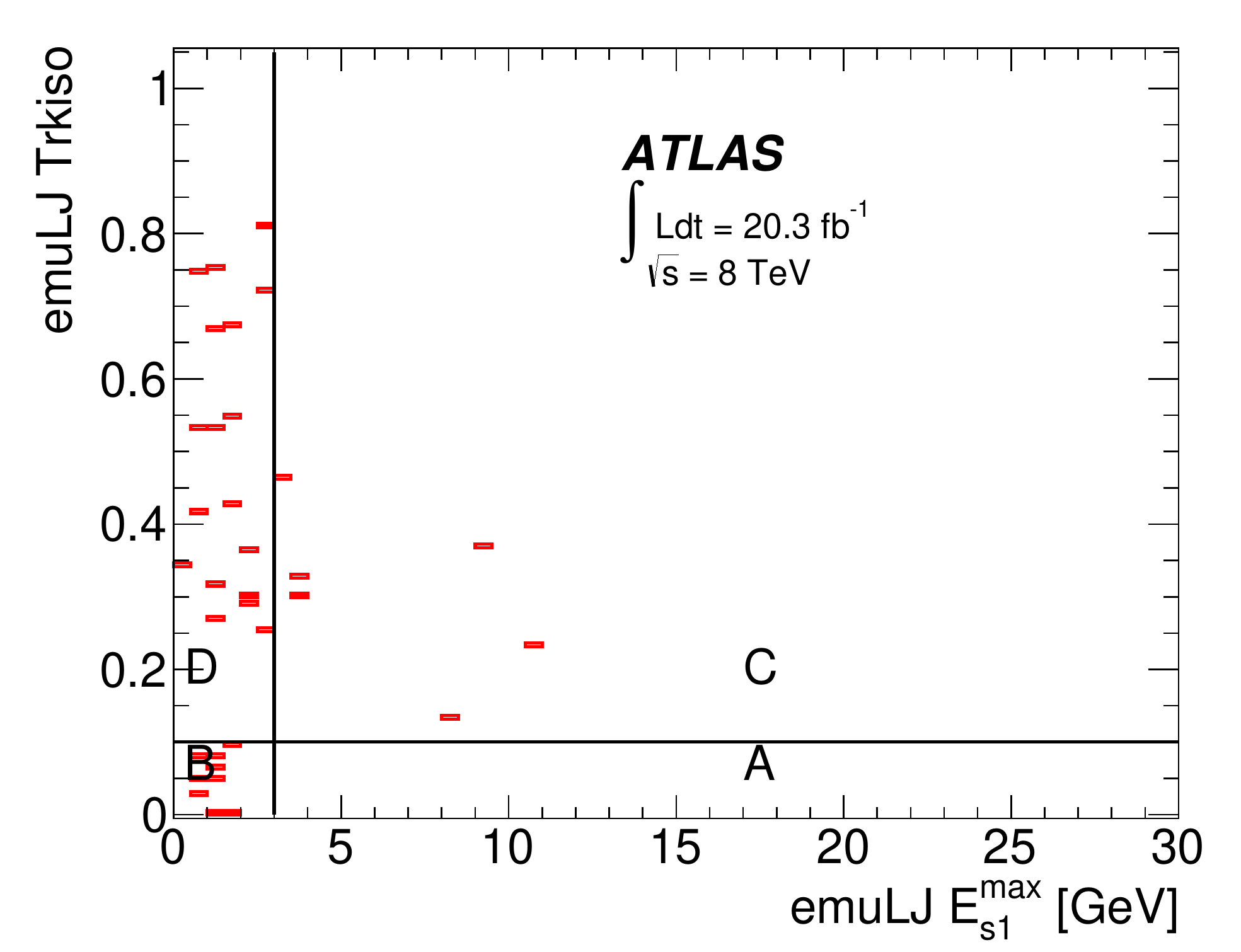}       
		\label{fig:EMuEMu_correlationSigStream}
          }
         \\
   	\caption{Scatter plot of the variables used in the ABCD-method for (a) eLJ--eLJ, (b) muLJ--muLJ, (c) eLJ--muLJ, (d) eLJ-- emuLJ, (e) muLJ--emuLJ, (f) emuLJ--emuLJ.  The horizontal and the vertical black lines indicate the cut values on both variables, as indicated in Table \ref{tab:cutvalues}.}
	\label{fig:correlation_SigStream}
	
\end{figure}

\noindent  In the case of the eLJ--eLJ channel the least correlated variables, $f_{\text{EM}}$ and $f_{\text{HT}}$, associated with the sub-leading eLJ are used for the ABCD plane after all other requirements are already applied. The two-dimensional distribution of these variables associated with the sub-leading eLJ is dominated by the multijet background in all four regions. Region $A$ is defined for events where the eLJ passes the $f_{\text{EM}}$ cut and its two leading tracks also pass the $f_{\text{HT}}$ cut; region $B$ for events where one or both leading tracks of that eLJ fail the $f_{\text{HT}}$ cut, but the eLJ passes the $f_{\text{EM}}$ requirement; region $C$ for events where both leading tracks pass the $f_{\text{HT}}$ cut, but the eLJ fails the $f_{\text{EM}}$ cut; region $D$ for events where that eLJ fails the $f_{\text{EM}}$ and one or both leading tracks fail the $f_{\text{HT}}$ cut. Similarly the B, C and D regions are shown in Figure \ref{fig:correlation_SigStream} for other LJ pairs. Table \ref{tab:abcdvar} summarizes the variables used for the ABCD-method for each channel.

\begin{table}[ht!]
\centering
\caption{List of nearly uncorrelated variables used in the ABCD-method for each channel.}
\begin{tabular}{| c | c | c |}
\hline
Channel & Variable 1 & Variable 2 \\
\hline
eLJ--eLJ       & sub-leading eLJ $f_{\text{EM}}$ & sub-leading eLJ $f_{\text{HT}}$ \\
muLJ--muLJ &  leading muLJ calorimeter isolation  & sub-leading muLJ calorimeter isolation \\
eLJ--muLJ   & eLJ $f_{\text{EM}}$  & muLJ calorimeter isolation \\
eLJ--emuLJ &  eLJ  $f_{\text{EM}}$ & emuLJ $E_{\text{T}}^{\text{had}}$ \\
muLJ--emuLJ &  muLJ calorimeter isolation & emuLJ track isolation \\
emuLJ--emuLJ &  emuLJ $E_{s1}^{\text{max}}$  & emuLJ track isolation \\
\hline
\end{tabular}
\label{tab:abcdvar}
\end{table}


Figure \ref{fig:correlation_SigStream} shows a two-dimensional plot of the ABCD variables from the observed data used for the\ LJ search. The $A$, $B$, $C$ and $D$  regions are determined by the cuts applied, as described in Section \ref{sec:eventsel}, to the two variables given in Table \ref{tab:abcdvar}. For visualization purposes, the 2D histogram shows only the leading track's $f_{\text{HT}}$ distribution for the eLJ--eLJ channel.\\

\section{Systematic uncertainties}
\label{sec:Syst}
The following effects are considered as possible sources of systematic uncertainty and are included as inputs to the likelihood.
\begin{itemize}

\item{\bf Luminosity} \\
The overall normalization uncertainty on the integrated luminosity is 2.8\%~\cite{lumi}.

\item{\bf Trigger} \\
The modelling of the lepton triggers is checked for events containing collimated electrons by comparing the efficiency for matching eLJs to the offline "medium" or "loose" \cite{Electron1} electrons using photo-conversion events in the data and MC samples as a function of $\Delta R$ between the two tracks with high threshold TRT hit fractions. The photo-conversion events are selected by a trigger which requires at least two photons with \et thresholds of 25 GeV and 35 GeV. The total systematic uncertainty of the signal efficiency due to the triggers used in the eLJ--eLJ channel is $13.5\%$. The systematic uncertainty on the efficiency of the dimuon triggers originates from two close muons which may fall in a single RoI and be identified as a single-muon at Level 1. The systematic uncertainty is determined as the difference between the ratio of data to simulation efficiencies and a straight line fit to the ratio vs. $\Delta R$ between two leading tracks in  a lepton-jet. The systematic uncertainty associated with the single-muon trigger is small ($ 0.6 \%$), as described in Ref. \cite{MuonTrig}. The systematic uncertainty associated with the multi-muon triggers is evaluated using a data-driven method applied to $J/\psi \to \mu^{+} \mu^{-}$ data and simulated samples. This uncertainty is $5.8 \%$ of the signal acceptance. 

The systematic uncertainty due to the triggers used in the mixed channel selection is evaluated as the weighted average uncertainty on the signal acceptance from the uncertainties associated with each of the relevant triggers. This uncertainty ranges from $3.3\%$ to $5.4\%$ depending on the $\gamma_d$ mass for the mixed channels having eLJ--muLJ and muLJ--emuLJ combinations. For other mixed channels (eLJ-- emuLJ and emuLJ--emuLJ) the uncertainty ranges from $3.1\%$ to $4.8\%$. For events where the triggers overlap, the largest of the two uncertainties is used.

\item{{\bf{ Lepton momentum resolution:}} The systematic uncertainty for the lepton momentum resolution is evaluated by smearing and shifting the momentum of the leptons by scale factors derived from comparison of  $Z \rightarrow \ell^{+}\ell^{-}$ in data and simulations, and by observing the effect of this shift on the signal efficiency. For electrons \cite{Electron1} the uncertainty is found to be less than $0.1 \%$, whereas for muons \cite{Muon1} it is found to be in the range $1.9\%$ to $5.2\%$ depending on the dark-photon mass value.}

\item{{\bf{Track reconstruction at small $\Delta R$:}} This systematic uncertainty is evaluated by studying differences in the track reconstruction efficiency between data and simulated samples. The efficiency is measured using a data-driven method for reconstructed $J/\psi \rightarrow \mu^+\mu^-$ candidates with a small $\Delta R$ (approximately  between $\Delta R$ values of 0.05 and 0.3)  between the two muons for data and simulated samples. The uncertainty due to this effect is found to be 8.4\%.}



\item{{\bf{Muon reconstruction at small $\Delta R$:}} The systematic uncertainty for the muon reconstruction in the muon spectrometer for small $\Delta R$  is described in Ref.~\cite{HiggsDisplacedLJ8TeVATLASpaper}. It is evaluated using a data-driven method using $J/\psi$ events to study the muon reconstruction efficiency estimation as a function of $\Delta R$ between two tracks, and the difference in efficiency between data and simulation is taken as a systematic uncertainty. The uncertainty is $5.4\%$.}

\item{{\bf{LJ variables:}}  In order to assess the size of the systematic uncertainty due to the mismodelling of LJ variables, the shapes of the discriminating variables are compared for $Z$ boson, photon conversions and multijet samples. Systematic uncertainties are assigned for each eLJ variable ($f_{\text{HT}}$, $E_{s1}^{\text{max}}$, $f_{s3}$, $f_{\text{EM}}$ and scaled track isolation) based on the cut efficiency difference of each of those variables between the data and MC samples containing $Z \rightarrow e^+e^-$ events. The validity of this procedure is checked by comparing the shapes of the lepton-jet variable distributions in the signal MC simulation of dark photons with the $Z \rightarrow e^+e^-$ MC simulation for various intervals of cluster \et and $\Delta R$ between two tracks in the lepton-jets. The electron cluster in the  $Z \rightarrow e^+e^-$ process is comparable to the cluster of two overlapping electrons from a dark photon in the longitudinal shower profiles, track isolation, and $f_{\text{HT}}$ variables. Since the emuLJ selection is based on variables very similar to those for the clusters in the EM calorimeter, the same sample of $Z \rightarrow e^+ e^-$ events comprising reconstructed eLJs is used to evaluate the systematic uncertainties on emuLJ, and cross-check the emuLJ variables' ($E_{s1}^{\text{max}}$, $f_{s3}$, $E_{\text{T}}^{\text{had}}$ and track isolation) distribution shapes in the multijet events from the data and MC simulation to see the impact of the presence of a muon in the LJ. The systematic uncertainties on the variables associated with the muLJ are obtained following a similar procedure to that for eLJ but instead using $J/\psi \rightarrow \mu^{+}\mu^{-}$ events, which are used given the lack of events for $Z \rightarrow \mu^{+}\mu^{-}$ where the muons should be within  $\Delta R = 0.5$. The systematic uncertainties for the six types of LJ pairs are given in Table~\ref{tab:SystematicsLJVariables}.}

\begin{table}[ht!]  
\footnotesize  
\centering
\caption{The relative systematic uncertainties associated with the signal acceptance due to the modelling of the discriminating variables in the six types of LJ pairs.}
\begin{tabular}{| c | c | c | c | c | c | c | c |}
    \hline
\cline{2-7} 
Variables    & eLJ--eLJ & muLJ--muLJ & eLJ--muLJ &  eLJ-- emuLJ & muLJ--emuLJ & emuLJ--emuLJ \\    
\hline
$f_{\text{HT}}$       		& 0.4 \%  &  -      &  0.2 \%  & 0.2 \%  & -       & -      \\    
Track isolation     	&  9.2 \% & 17.0 \% &  13.0 \% & 9.2 \% & 13.0 \% & 9.2 \% \\    
$E_{s1}^{\text{max}}$     	& 0.2 \%  &  -      &  0.1 \%  & 5.5 \%  & 5.4 \%  & 11.1 \% \\
$f_{s3}$            	& 12.9 \%  &  -      &  6.0 \%  & 9.7 \%  & 3.3 \%  & 6.8 \% \\
$E_{\text{T}}^{\text{had}}$       	& -       &  -      &  -       & 0.1 \%  & 0.1 \%  & 0.2 \% \\
$f_{\text{EM}}$            	& 0.9 \%  &  -      &  0.4 \%  & 0.4 \%  & -       & -      \\
Muon calorimeter isolation & -       & 6.7 \%  &  3.3 \%  & -       & 3.3 \%  & -      \\
\hline
\end{tabular}     
\label{tab:SystematicsLJVariables}
\end{table}

\item{{\bf{Backgrounds:}} The two discriminating variables used in the ABCD data-driven background estimation show a small correlation of about 6\% in all LJ pairs except emuLJ--emuLJ. The effect of this correlation is incorportated in the background estimation using the ABCD-likelihood method. This is done by introducing a nuisance parameter in the likelihood fit. The correlation is taken into account as a systematic uncertainty on the ABCD-likelihood estimation for the background originating from jets in the multijet and $W$/$Z$/$\gamma$/$\gamma^{*}$ + jets processes. An additional $4.5 \%$ uncertainty is assigned to the background estimate due to a small pile-up dependence of the track isolation variable (multijet background in region $A$ could suffer from pile-up effects).}

\end{itemize}

\clearpage

\section{Observed events in data and background estimation}
\label{sec:results}

\noindent The number of observed events in data for all six channels in signal region $A$ is shown in Table~\ref{tab:Data}. The expected yields for background processes estimated from the data-driven ABCD-likelihood as well as  the total background estimated in region $A$  are also shown in this table. The difference between the number of events estimated with the ABCD-likelihood method and the total background comes from diboson and $t\bar{t}$ contributions. The expected number of events in the MC simulation is shown in the Appendix (Tables \ref{tab:sig1} and \ref{tab:sig2}). The numbers shown for the background are for a dark-photon mass of 300 MeV and two dark-photon production in the SUSY-portal model. The variation of the total background estimate from changing the dark-photon multiplicity, the dark-photon mass or even considering a different model (Higgs-portal) is found to be less than 4\%. The data are found to be in good agreement with the background prediction.\\

\begin{table}[ht!] 
\centering
\caption{Number of signal and background events in signal region $A$. The expected yields for background processes estimated from the data-driven ABCD-likelihood as well as the total background estimation taking into account diboson and $t\bar{t}$ contributions are also shown.}
\begin{tabular}{| c | c | c | c | c | c | c | c |}
    \hline
 {Channel} & Background (ABCD-likelihood method) & Background (total)& Observed events in data\\
     \hline
     eLJ--eLJ  & 2.9 $\pm$ 0.9 & 4.4 $\pm$ 1.3  & 6 \\
     muLJ--muLJ  &  2.9 $\pm$ 0.6  &4.4 $\pm$ 1.1  & 4 \\
    eLJ--muLJ   &  6.7 $\pm$ 1.4  & 7.1 $\pm$ 1.4 & 2 \\
    eLJ--emuLJ &  7.8 $\pm$ 2.0  &  7.8 $\pm$ 2.0 & 5 \\
    muLJ--emuLJ & 20.2 $\pm$ 4.5 &  20.3 $\pm$ 4.5 & 14 \\ 
    emuLJ--emuLJ &  1.3 $\pm$ 0.8 &  1.9 $\pm$ 0.9 & 0  \\
    \hline
     \end{tabular}
\label{tab:Data}
\end{table}

\section{Interpretation and limits}
\label{sec:limits}

No significant deviation from SM predictions is found, and 95\% confidence-level upper limits are placed on the contribution of new phenomena beyond the SM on the number of events with lepton-jets. A likelihood-based approach is employed for hypothesis-testing and limit calculation, using the CL(s) technique \cite{CLs}.

\noindent All systematic uncertainties discussed in Section~\ref{sec:Syst} are taken into account. The expectation and uncertainties are calculated using the HistFactory statistical tool \cite{Cramer}. A log-likelihood ratio (LLR) is used as the test statistic, defined as the ratio of the signal-plus-background hypothesis to the background-only hypothesis. Ensembles of pseudo-experiments were generated for the signal-only hypothesis and the signal+background hypothesis, varying the LLR according to the statistical and systematic uncertainties. For a given hypothesis, the combined likelihood is the product of the likelihoods for the channels considered, each resulting from the product of a Poisson distribution representing the statistical fluctuations of the expected total event yield, and of Gaussian distributions representing the effect of the systematic uncertainties. The upper limits were determined by performing a scan of $p$-values corresponding to LLR values larger than the one observed in data. Limits are placed for squark + squark $\rightarrow 2 \gamma_d + \text{X}$, squark + squark $\rightarrow 2 (s_{d} \rightarrow \gamma_d \gamma_d) + \text{X}$, $H \rightarrow 2 \gamma_d +\text{X}$ and $ H \rightarrow  2 (s_{d} \rightarrow \gamma_d \gamma_d) + \text{X}$ processes in the electron, muon and mixed channels. \\

\noindent The 95\% confidence-level upper limits on signal are expressed in terms of the cross section times BR for the production of two lepton-jets, which are shown in the Appendix (Tables~\ref{tab:eLJeLJlimnum}--\ref{tab:emuLJemuLJlimnum}) and are based on the following formula:

\begin{equation}
\sigma \times BR = \dfrac{N_{\text{limit}}}{\mathcal{L} \times \epsilon_{\text{s}}}
 \label{eqn:xsecBR}
\end{equation}

\noindent  Here $N_{\text{limit}}$ is the upper limit on the signal yield, $\mathcal{L}$ represents the integrated luminosity, and $\epsilon_{\text{s}}$ represents the signal efficiency including the trigger efficiency, signal acceptance, the reconstruction efficiency for two LJs and the efficiency of the selection criteria for the discriminating variables on both LJs. The uncertainties on the luminosity and on the efficiencies are taken into account in the likelihood that derives $N_{\text{limit}}$.\\

\begin{figure}[ht!]
\includegraphics[width=0.45\textwidth]{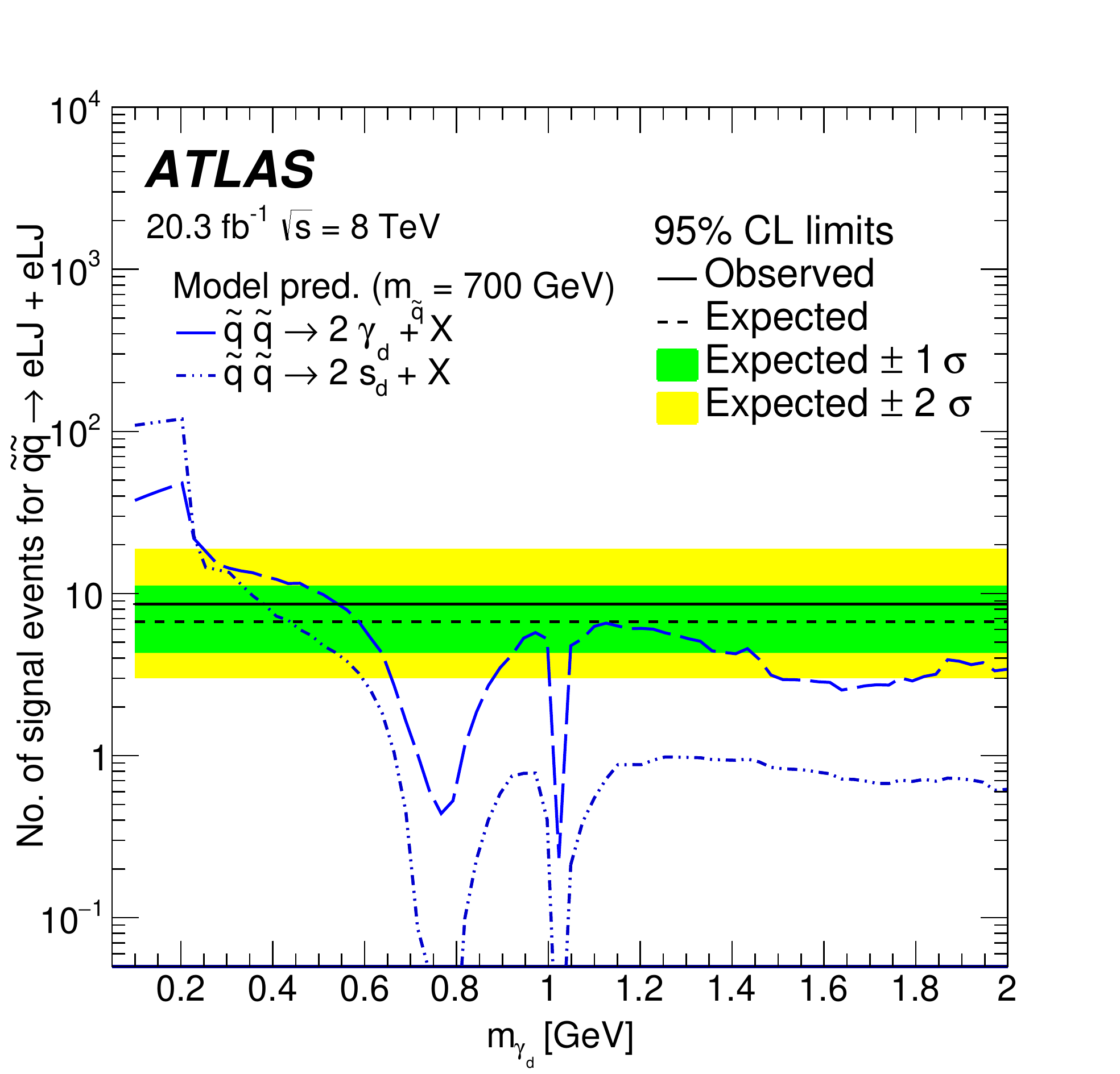}
\includegraphics[width=0.45\textwidth]{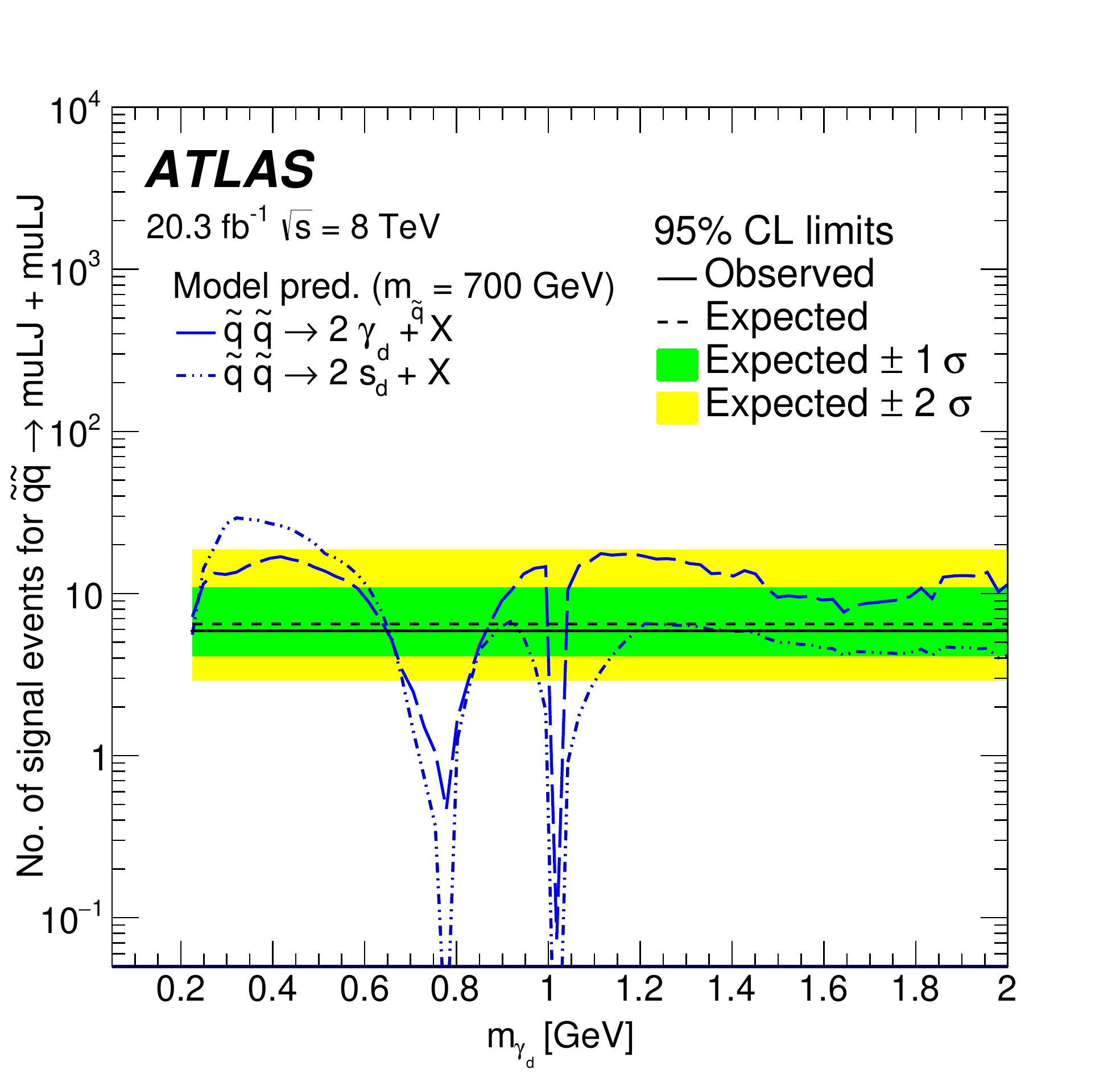}
\includegraphics[width=0.45\textwidth]{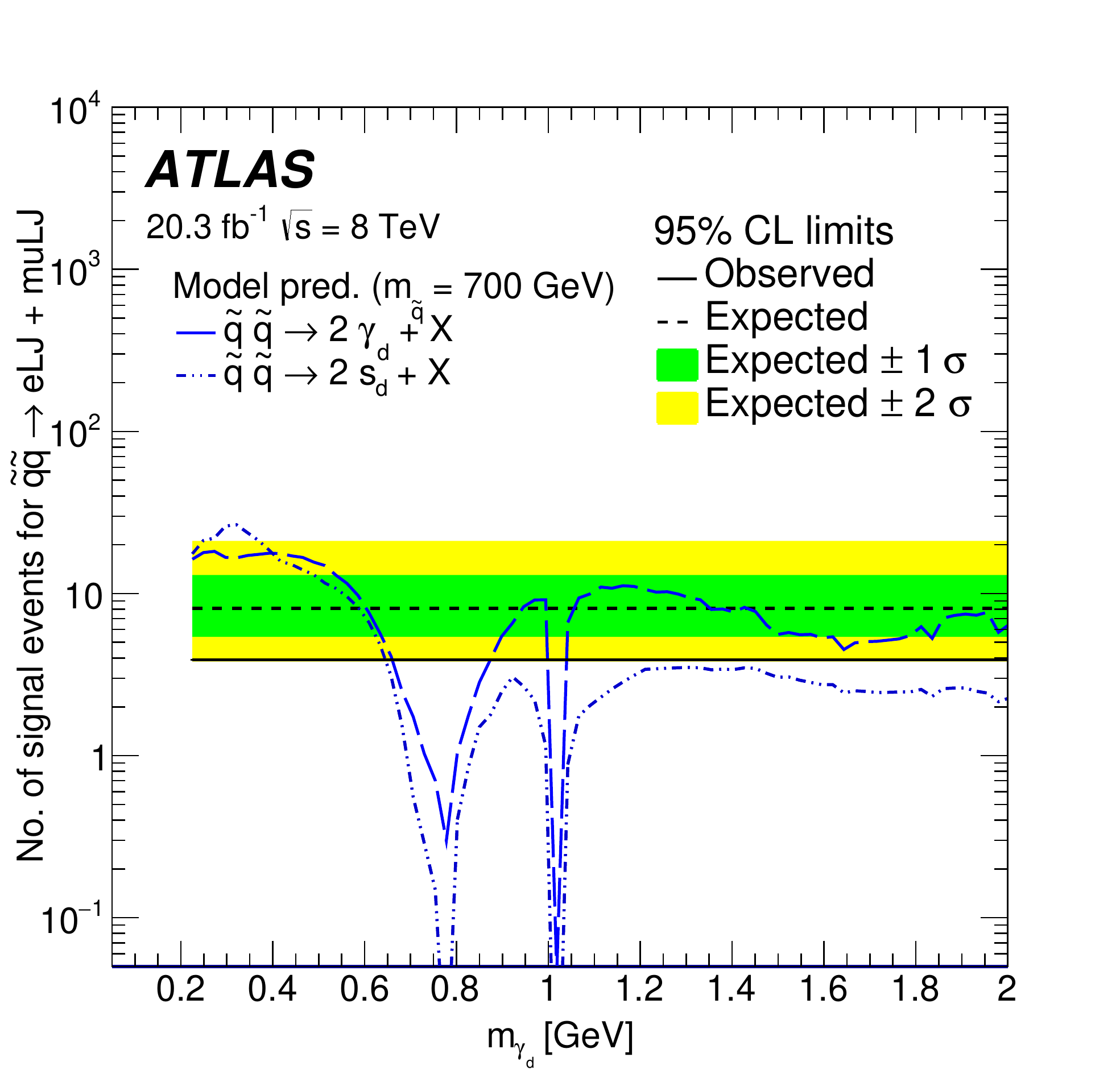}
\includegraphics[width=0.45\textwidth]{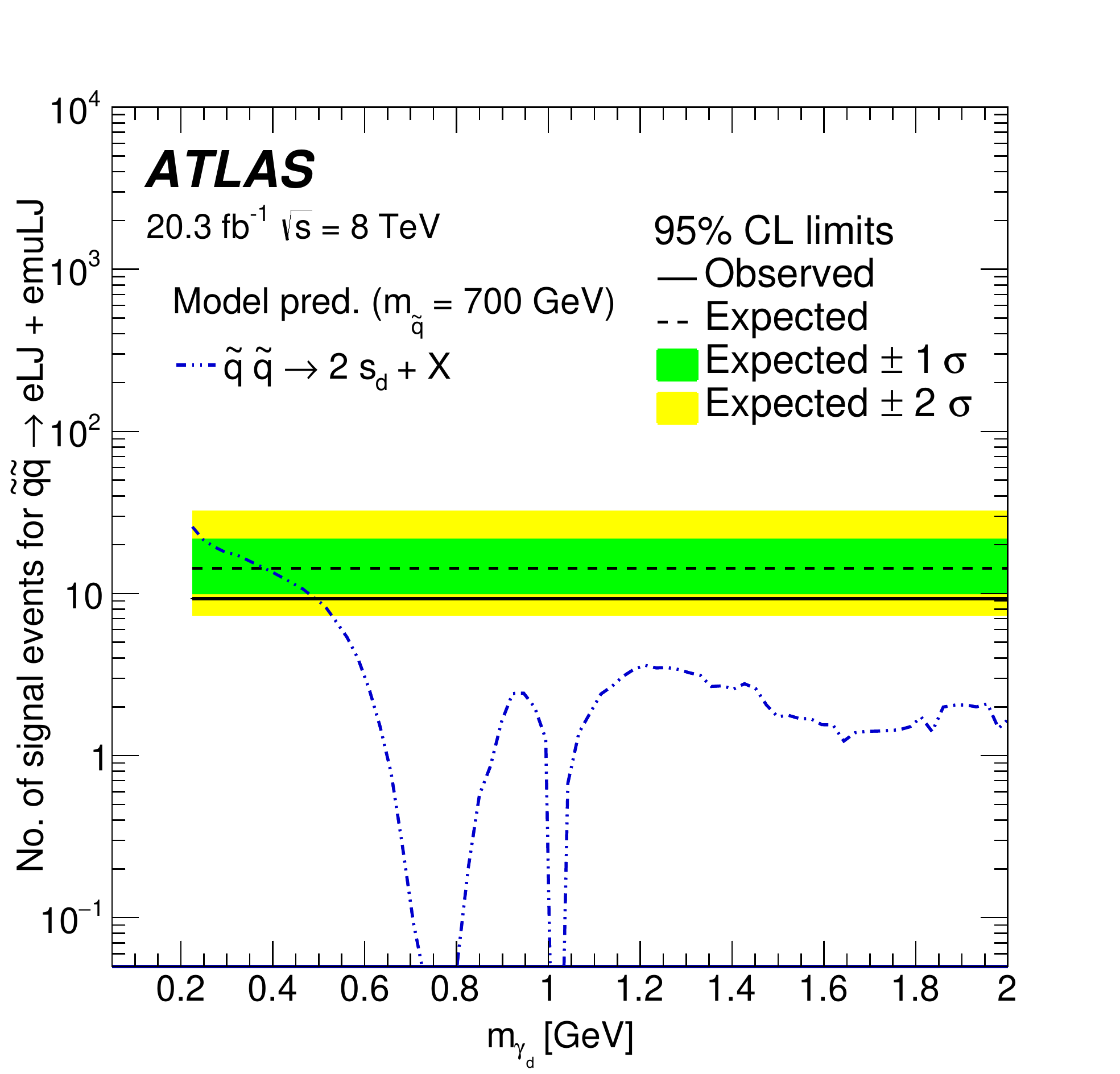}
\includegraphics[width=0.45\textwidth]{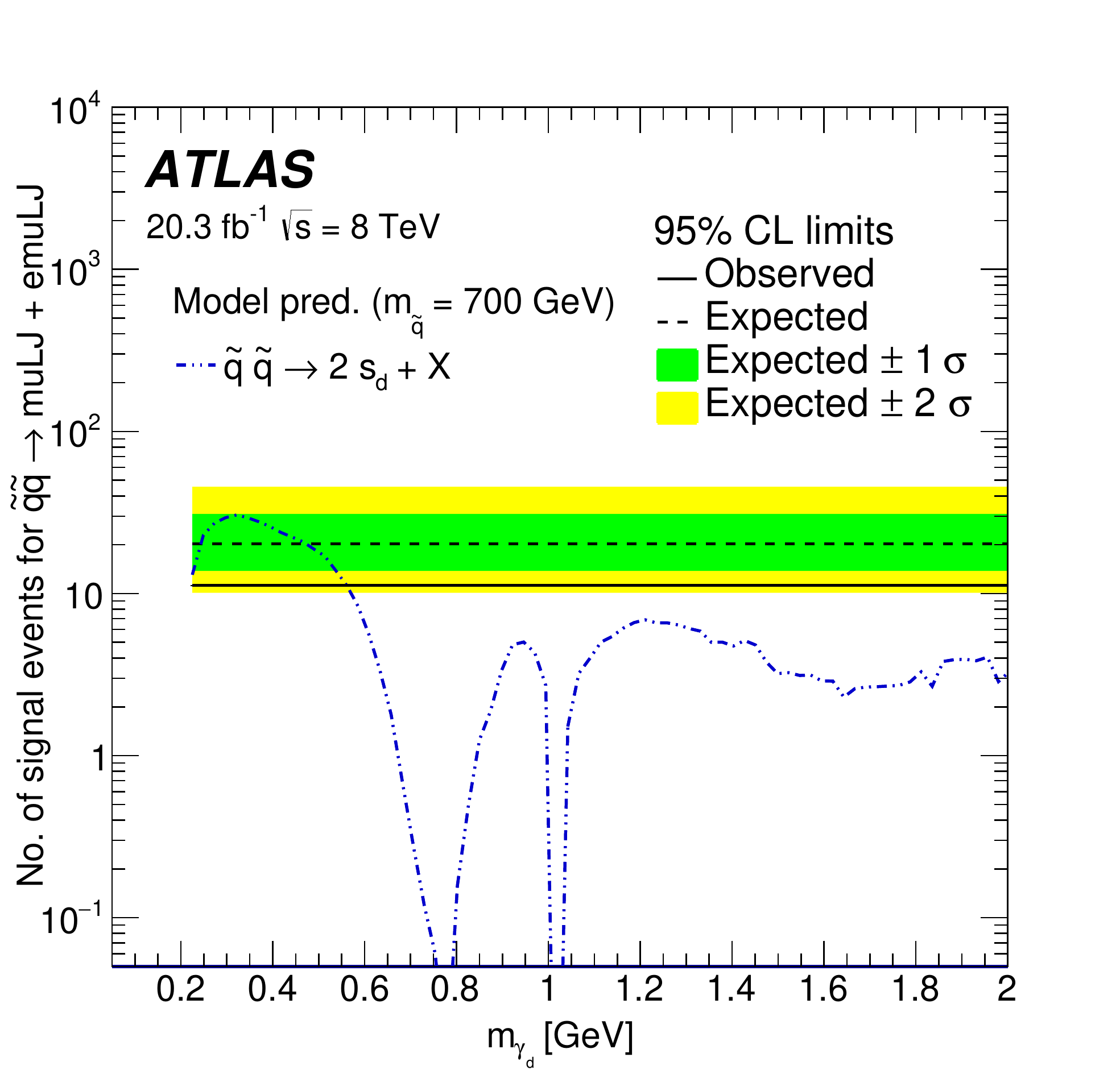}
\includegraphics[width=0.45\textwidth]{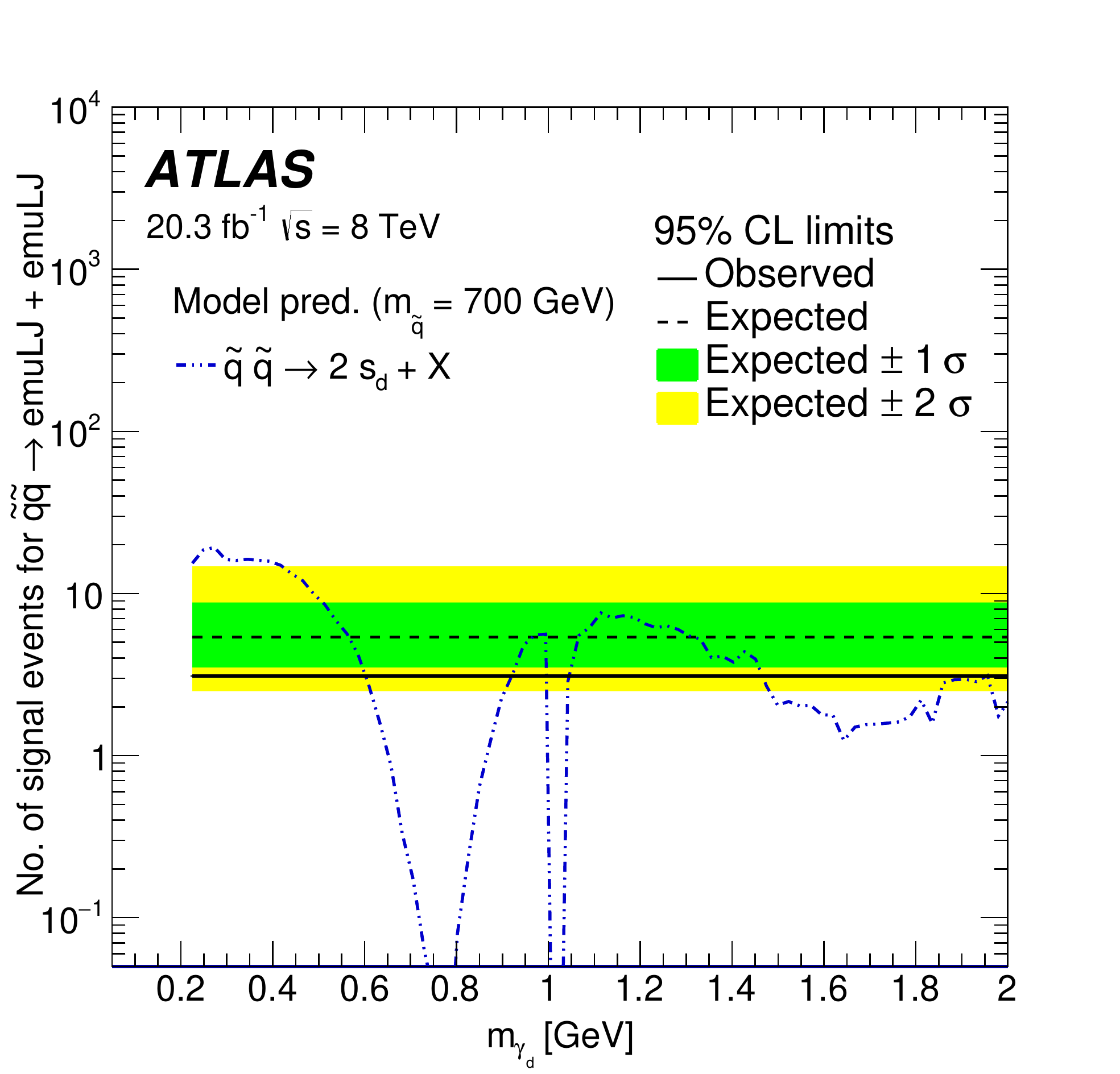}
\caption{The $95\%$ confidence-level observed and expected upper limits on the number of signal events for final-states consisting of two lepton-jets in the {\bf{eLJ--eLJ}}, {\bf{muLJ--muLJ}}, {\bf{eLJ--muLJ}}, {\bf{eLJ--emuLJ}}, {\bf{muLJ--emuLJ}} and {\bf{emuLJ--emuLJ}} channels. Results based on \intlumi of integrated luminosity are shown in these figures. The model predictions for the production of 2 \gammad + X and $2 (s_{d} \rightarrow \gamma_d \gamma_d)$ + X via SUSY-portal topologies for various $\gamma_d$ mass  values are also overlaid.} 
\label{fig:Nlim}
\end{figure}
\begin{figure}[ht!]
\begin{center}
\includegraphics[width=0.5\textwidth]{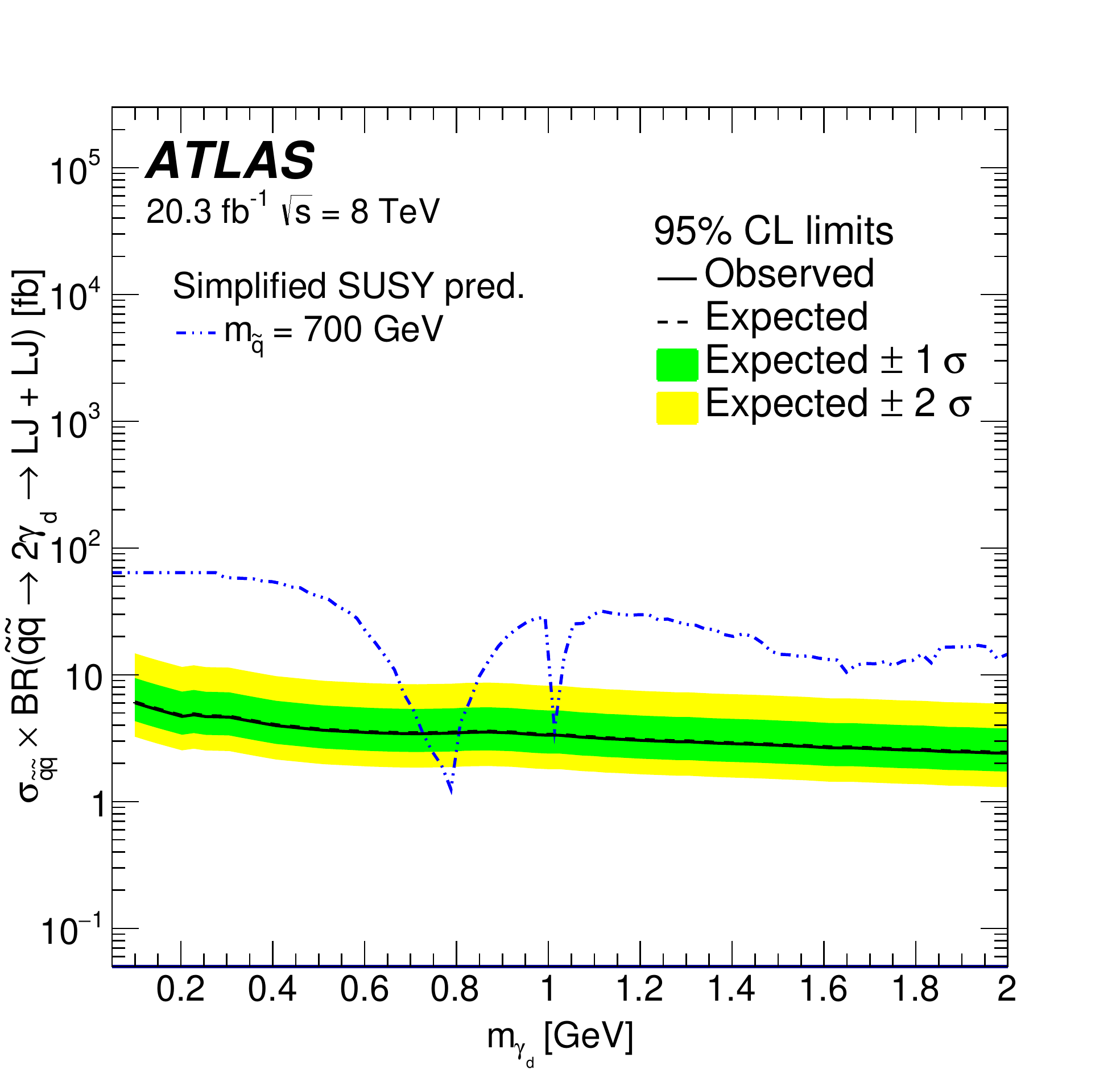}
\caption{The $95\%$ CL observed and expected upper limits on the cross section times branching ratio into final-states with two lepton-jets in the case of $2 \gamma_{d}$ + X production via the SUSY-portal topology based on the combined results of the {\bf{eLJ--eLJ}, \bf{muLJ--muLJ}, \bf{eLJ--muLJ}} channels. The limit is plotted as a function of dark-photon mass $m_{\gamma_d}$, and changes with mass due to a small dependence of signal efficiency on the $\gamma_d$ mass.}
\label{fig:LJLJlim}
\end{center}
\end{figure}

\noindent The expected and observed limits on the number of signal events are shown in Figure~\ref{fig:Nlim} for all six channels.  The theoretical predictions for the signal in the $2 \gamma_d + X$ and $4 \gamma_d$ + X final-states for the SUSY-portal model are also illustrated in this figure. At $m_{\gamma_d}$ = 0.7 GeV, the branching fraction of dark photons into lepton pairs is around 15\% due to higher decay probabilities into $\rho$ and $\phi$ mesons,  which is the reason why the signal expectation is small for this mass point. Figure~\ref{fig:LJLJlim} shows the 95\% CL combined upper limit on the cross section times branching ratio for the $2 \gamma_d$ + X topology.  \\

\begin{figure}[ht!]
\begin{center}
\includegraphics[width=0.5\textwidth]{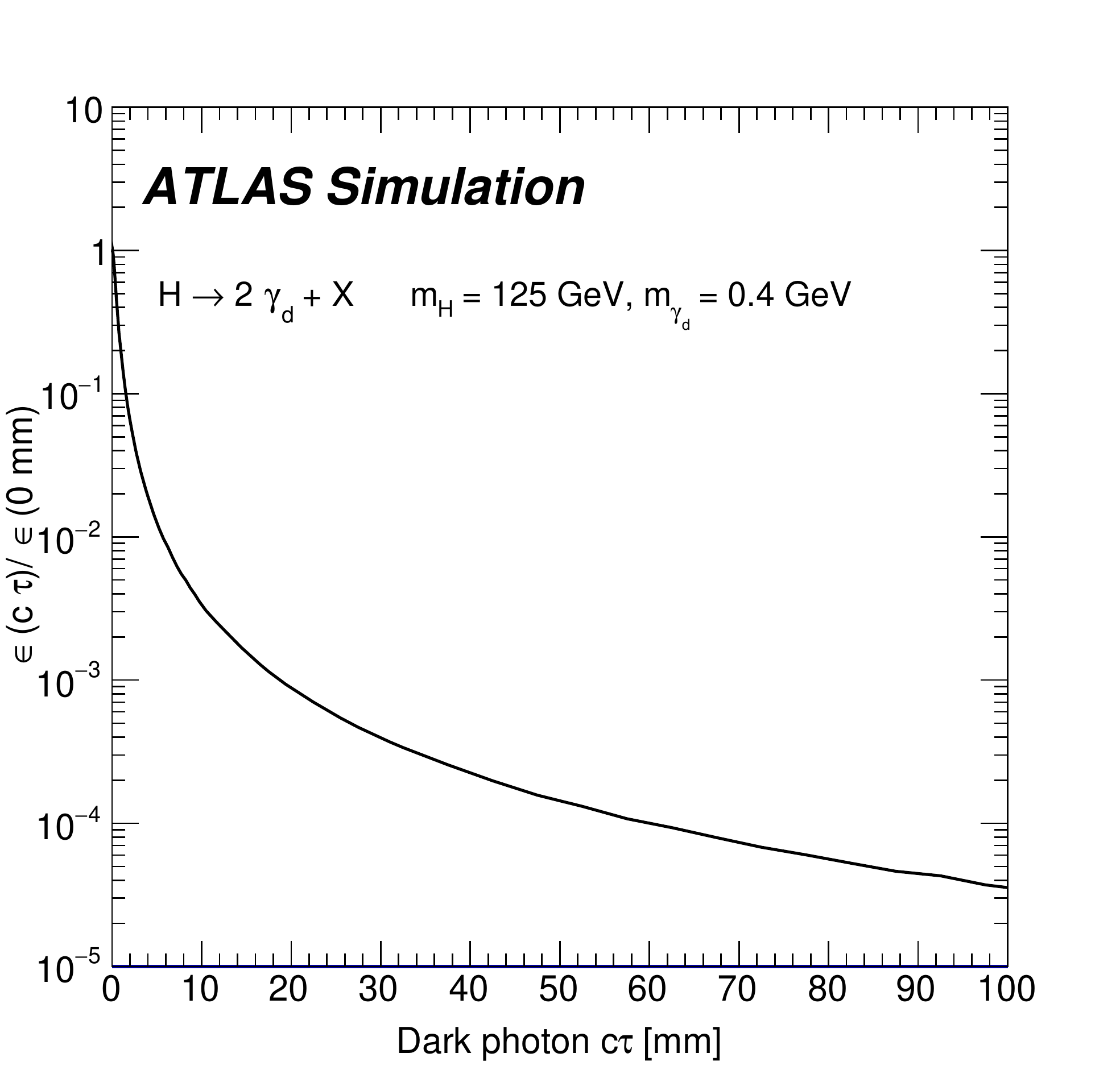}
\caption{The detection efficiency ratio with respect to the reference model efficiency with zero mean life time is plotted as a function of $\gamma_d$ mean lifetime $c\tau$. The extrapolated efficiency for other $c\tau$ is estimated from the $H \rightarrow 2 \gamma_d + \text{X}$ MC sample generated with $c \tau = 47$ mm by studying the efficiency as a function of $\gamma_d$ decay position. A 45\% uncertainty is associated on this efficiency scaling as described in the text.}
\label{fig:effratiovsctauFRVZmodel}
\end{center}
\end{figure}

\noindent The ratio of the average efficiency at a given $c\tau$ to the efficiency at $c\tau =0$ mm is used to rescale the expected number of signal events estimated from the reference $H \rightarrow 2 \gamma_d + \text{X}$ sample. Here $c\tau$ is the mean lifetime of the dark photon used in the simulation and the efficiency is the average efficiency for this mean lifetime. This efficiency scaling is shown in Figure~\ref{fig:effratiovsctauFRVZmodel}. A  large number of pseudo experiments is generated for $c\tau$ ranging from 0 - 100 mm. The average efficiency ratios are obtained based on the lepton-jet efficiency dependence on the proper decay length.The average efficiency curves are valid for all channels. This is due to the B-layer hit requirements of at least two tracks in the lepton-jets, which makes the lepton-jet reconstruction efficiency vanish for all lepton-jets types with the proper decay length > 52 mm.  Therefore there is no further impact due to the lepton-jet reconstruction, their selection cuts and the triggers requirements.    \\

\begin{figure}[ht!]
\begin{center}
\includegraphics[width=0.5\textwidth]{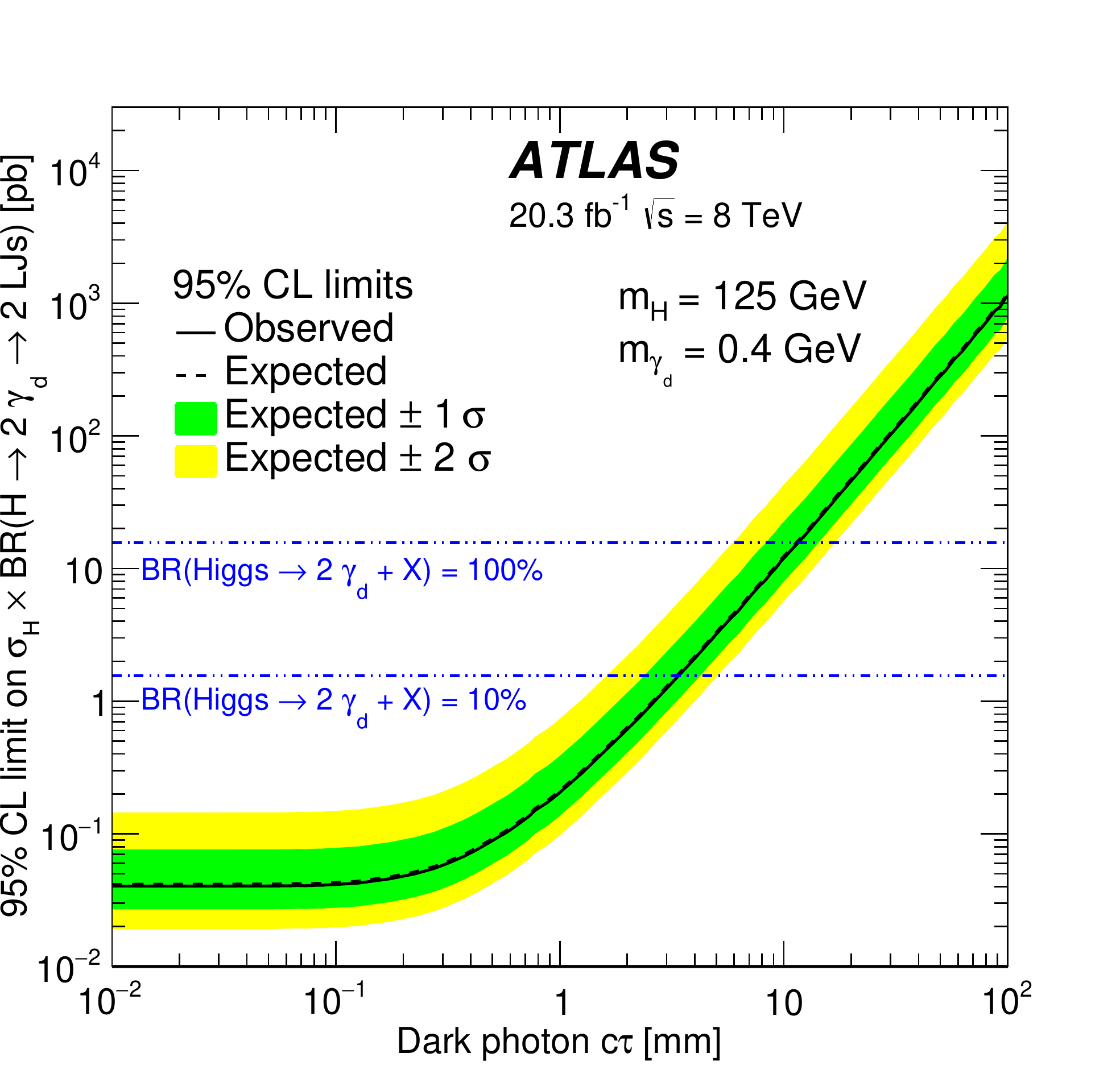}
\caption{The $95\%$ confidence-level observed and expected upper limits on the cross section times branching fractions into final-states consisting of two lepton-jets in the production of $2 \gamma_{d}$ + X via Higgs-portal topology for $m_{\gamma_d} = 0.4$ GeV based on the combined results of the {\bf{eLJ--eLJ}, \bf{muLJ--muLJ}, \bf{eLJ--muLJ}} channels. The limit is drawn as a function of lifetime $c \tau$. The results for various lifetimes ranging up to $100$ mm are derived by extrapolating the detection efficiency using the curve as described in Figure~\ref{fig:effratiovsctauFRVZmodel}. The comparison with the theoretical prediction (dashed line) for $10\%$ BR of Higgs boson decay to two dark photons shows that values of $c \tau$ below $3.2$ mm are excluded at 95$\%$ confidence-level.}
\label{fig:sigmaBR0p95limitvsctauFRVZmodel}
\end{center}
\end{figure}

\noindent The $95\%$ confidence-level upper limit on the production cross section times BR to two lepton-jets in the $H \rightarrow 2 \gamma_d + \text{X}$ model is obtained as a function of lifetime $c\tau$ as shown in Figure~\ref{fig:sigmaBR0p95limitvsctauFRVZmodel}, after taking into account the uncertainty associated with the efficiency scaling for $c\tau$. A $45\%$ uncertainty is assigned on the efficiency scaling based on a comparison of the extrapolated signal expectation at 47 mm $c\tau$  with the direct estimate of the expected signal for the dark-photon simulation sample generated with  $c\tau$  = 47 mm. The extrapolated signal expectation is obtained by scaling the signal expectation for the dark-photon sample with $c\tau$ = 0 mm by the average efficiency ratio at $c\tau$ = 47 mm as given by the curve in  Figure~\ref{fig:sigmaBR0p95limitvsctauFRVZmodel}.  The decays of the dark photons into leptons are simulated with an exponential decay law. The limit is based on combined results from the eLJ--eLJ, muLJ--muLJ and eLJ--muLJ channels. The emuLJ channels are not used as they do not contribute to the $H \rightarrow 2 \gamma_d + \text{X}$ topology, given that a single $ \gamma_d $ cannot decay into a pair of leptons of different flavour. The comparison with the theoretical prediction (dashed line) for $10\%$ BR of Higgs boson decay to two dark photons \footnote{This is an arbitrary choice, as the BR of the Higgs boson to dark photons is not theoretically known.} shows that values of $c \tau$ below $3.2$ mm are excluded at 95$\%$ confidence-level.\\

\noindent The results are also interpreted as a two-dimensional exclusion contour in the plane of the kinetic mixing parameter $\epsilon$ and the $\gamma_d$ mass for the $H \rightarrow 2 \gamma_d + \text{X}$ topology. As only one mass benchmark ($m_{\gamma_d}$ = 0.4 GeV) is generated for the $H \rightarrow 2 \gamma_d + \text{X}$ topology, the signal efficiencies are derived from the MC sample for the Higgs-portal in the eLJ--eLJ, muLJ--muLJ and eLJ--muLJ channels for that benchmark. For other $\gamma_d$ mass benchmarks, the extrapolation corrections for all $\gamma_d$ masses are obtained using the SUSY-portal MC samples. These correction factors are then used to rescale the efficiency at 0.4 GeV in the $H \rightarrow 2 \gamma_d + \text{X}$ topology to derive efficiencies for other $\gamma_d$ masses in the [0.1--2.0] GeV interval. As the efficiency dependence is found to be small for the SUSY samples across various dark-photon mass points in the $2 \gamma_d + \text{X}$ topology, it is assumed for the Higgs-portal model that the efficiencies scale the same way as for the SUSY-portal samples with respect to 0.4 GeV dark-photon sample. Based on the variations in efficiencies across different dark-photon masses with respect to 0.4 GeV dark-photon sample, the following uncertainties are assigned to the efficiencies: 60\%  on the eLJ--eLJ channel, 200\% on the muLJ--muLJ and 30\% on the eLJ--muLJ channel. \\

\noindent In order to allow a comparison with the displaced lepton-jets analysis \cite{HiggsDisplacedLJ8TeVATLASpaper}, $90\%$ confidence-level exclusion limits are derived for $H \rightarrow 2 \gamma_d + \text{X}$ production for various ($5\%$, $10\%$, $20\%$, $40\%$) branching fractions by combining the results from the eLJ--eLJ, muLJ--muLJ and eLJ--muLJ channels after taking into account all systematic uncertainties. 

\begin{figure}[ht!]
\begin{center}
\includegraphics[width=0.7\textwidth]{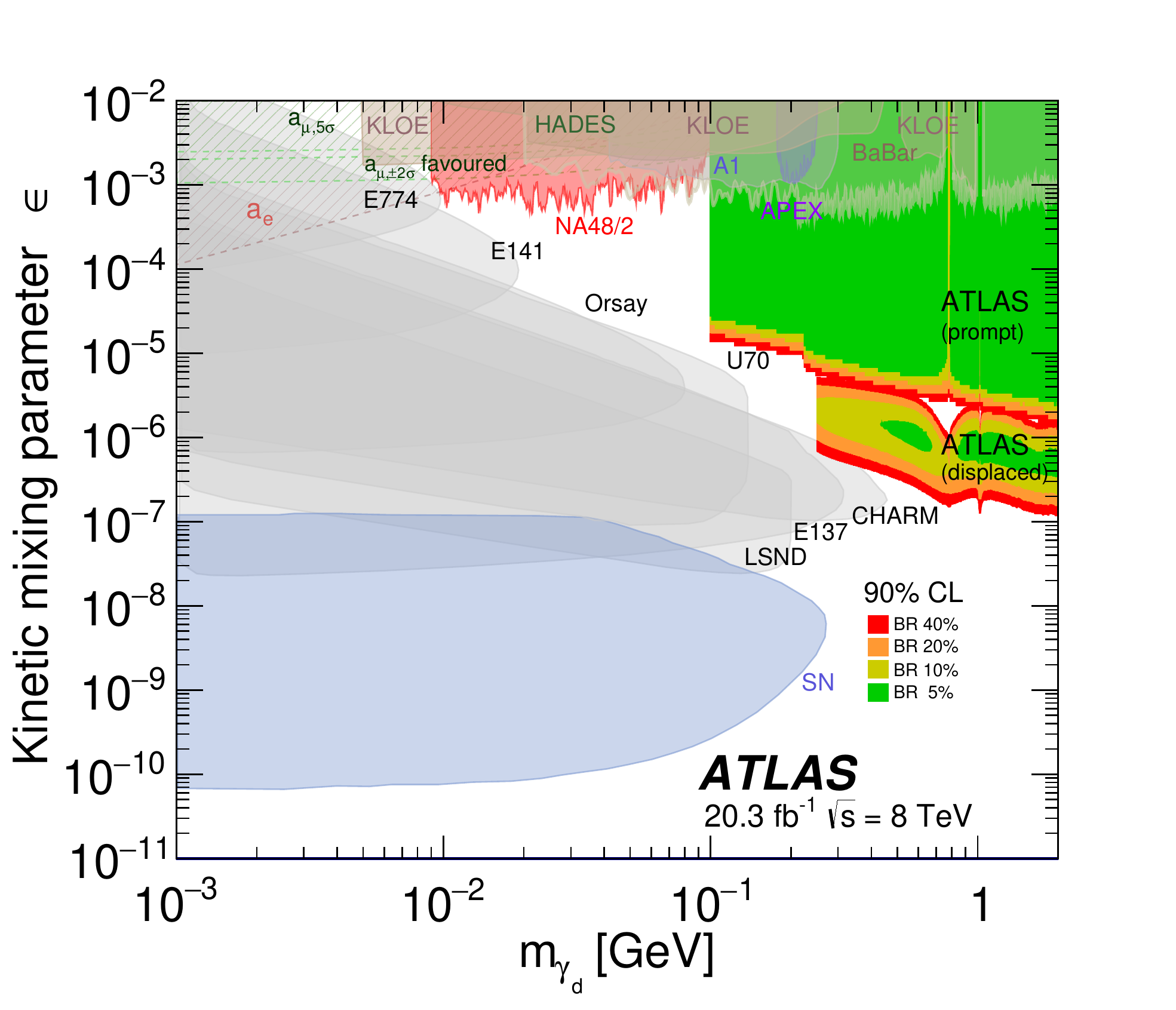}
\caption{A two-dimensional exclusion plot in the dark-photon mass $m_{\gamma_d}$ and the kinetic mixing $\epsilon$ parameter space, taken from Ref.~\cite{ExclusionContPlot}.  The branching ratios are for $H \rightarrow 2 \gamma_d + \text{X}$ decays.  The $90\%$ confidence-level exclusion region for prompt $H \rightarrow 2 \gamma_d + \text{X}$ production with $5\%$, $10\%$, $20\%$ and $40\%$ branching fractions into $2$ \gammad + X decay are extracted based on the present combined results from the eLJ--eLJ, muLJ--muLJ and eLJ--muLJ channels using 8 TeV data at ATLAS. The excluded regions are derived based on comparing the total signal expectation for the $H \rightarrow 2 \gamma_d + \text{X}$ production model with the $90 \%$ confidence-level upper limit on the signal expectation. The legend represents the ATLAS exclusions by both the prompt (this paper) and displaced lepton-jet analyses \cite{HiggsDisplacedLJ8TeVATLASpaper}. }
\label{fig:epsilonVsmass}
\end{center}
\end{figure}

Figure \ref{fig:epsilonVsmass} shows the 90\% confidence-level exclusion contour interpreted in the $\epsilon$ and $\gamma_d$ mass plane in the $\epsilon$ region $10^{-2}$--$10^{-6}$ and in the mass region [0.1--2] GeV for 5\%, 10\%,  20\%, 40\% branching fractions of the Higgs boson decay to $2 \gamma_d$ + X. In the low-mass region, below the $\mu^{+} \mu^{-}$ threshold, only the results from the eLJ--eLJ channel contribute. The results shown in the figure depend on the coupling of the dark photon to the SM photon $\epsilon$, and on the mass of the dark photon. The figure also shows other excluded regions from a search for non-prompt lepton-jets at ATLAS~\cite{HiggsDisplacedLJ8TeVATLASpaper} and from other experiments. Shown are existing $90\%$ confidence-level exclusion regions from beam-dump experiments E137, E141, and E774~\cite{beam-dump1a,beam-dump2,beam-dump3}, Orsay~\cite{beam-dump0}, U70~\cite{beam-dump1}, CHARM~\cite{fixed-target6}, LSND~\cite{fixed-target5}, A1~\cite{fixed-target2}, the electron and muon anomalous magnetic moment~\cite{emu-magneticmoment0,emu-magneticmoment1,emu-magneticmoment2}, HADES~\cite{fixed-target4}, KLOE~\cite{epluseminus-colliders1, Babusci1, Babusci2, Anastasi, epluseminus-colliders2}, the test-run results reported by APEX~\cite{fixed-target1}, an estimate using BaBar results~\cite{DecayLengthEps,epluseminus-colliders2,epluseminus-colliders4}, and constraints from astrophysical observations~\cite{astrophysexp1,astrophysexp2}. While other experiment's results are independent of the topology of the dark-photon production, the ATLAS results depend on the topology, i.e Higgs boson mass and its production mechanism, and its decay into dark photons. The results from CMS are shown elsewhere \cite{CMS_muons}.
The $\epsilon$ values are evaluated using Equation (\ref{eqn:gamma}) combined with R-ratio measurements of $e^{+}e^{-}$ collider data \cite{epluseminus-colliders1, Babusci1, Babusci2, Anastasi, epluseminus-colliders2,epluseminus-colliders4}, where the R-ratio is the ratio of the hadronic cross section to the muon cross section in electron–positron collisions. The expression for $\epsilon$ with respect to these variables is

\begin{eqnarray}
\epsilon^{2} = \frac{\hbar c}{c\tau (\Gamma_{e^{+}e^{-}} + \Gamma_{\mu^{+}\mu^{-}} + R \Gamma_{\mu^{+}\mu^{-}} ) } ,
\end{eqnarray}    

where $\hbar$ is the reduced Planck constant, $R$ is the R-ratio, while $\Gamma_{e^{+}e^{-}}$ and $\Gamma_{\mu^{+}\mu^{-}}$ are given by  Equation (\ref{eqn:gamma}).    

%
\section{Conclusions}
\label{conclusion} 
A search for prompt lepton-jets using 20.3 fb$^{-1}$ of $pp$ collision data collected with the ATLAS detector at the LHC at a centre-of-mass energy of 8 TeV is performed. Such lepton-jets are expected from decays of low-mass dark photons in some model extensions to the SM. The analysis requires events with at least two lepton-jets. \\

\noindent No significant excess of events compared to the SM expectation is observed in any of the analysed channels, and 95\% confidence-level upper limits are computed on the production cross section times branching ratio for two prompt lepton-jets in SUSY-portal and Higgs-portal models. The results are also interpreted in terms of a 90\% confidence-level exclusion region in kinetic mixing and dark-photon mass parameter space, based on combined results from the $H \rightarrow 2 \gamma_d + \text{X}$ topology. These results provide exclusion in regions of parameter space previously unexplored and extend the results of other searches.

%

\clearpage
\section*{Acknowledgements}

We thank CERN for the very successful operation of the LHC, as well as the
support staff from our institutions without whom ATLAS could not be
operated efficiently.

We acknowledge the support of ANPCyT, Argentina; YerPhI, Armenia; ARC, Australia; BMWFW and FWF, Austria; ANAS, Azerbaijan; SSTC, Belarus; CNPq and FAPESP, Brazil; NSERC, NRC and CFI, Canada; CERN; CONICYT, Chile; CAS, MOST and NSFC, China; COLCIENCIAS, Colombia; MSMT CR, MPO CR and VSC CR, Czech Republic; DNRF, DNSRC and Lundbeck Foundation, Denmark; IN2P3-CNRS, CEA-DSM/IRFU, France; GNSF, Georgia; BMBF, HGF, and MPG, Germany; GSRT, Greece; RGC, Hong Kong SAR, China; ISF, I-CORE and Benoziyo Center, Israel; INFN, Italy; MEXT and JSPS, Japan; CNRST, Morocco; FOM and NWO, Netherlands; RCN, Norway; MNiSW and NCN, Poland; FCT, Portugal; MNE/IFA, Romania; MES of Russia and NRC KI, Russian Federation; JINR; MESTD, Serbia; MSSR, Slovakia; ARRS and MIZ\v{S}, Slovenia; DST/NRF, South Africa; MINECO, Spain; SRC and Wallenberg Foundation, Sweden; SERI, SNSF and Cantons of Bern and Geneva, Switzerland; MOST, Taiwan; TAEK, Turkey; STFC, United Kingdom; DOE and NSF, United States of America. In addition, individual groups and members have received support from BCKDF, the Canada Council, CANARIE, CRC, Compute Canada, FQRNT, and the Ontario Innovation Trust, Canada; EPLANET, ERC, FP7, Horizon 2020 and Marie Skłodowska-Curie Actions, European Union; Investissements d'Avenir Labex and Idex, ANR, Region Auvergne and Fondation Partager le Savoir, France; DFG and AvH Foundation, Germany; Herakleitos, Thales and Aristeia programmes co-financed by EU-ESF and the Greek NSRF; BSF, GIF and Minerva, Israel; BRF, Norway; the Royal Society and Leverhulme Trust, United Kingdom.

The crucial computing support from all WLCG partners is acknowledged
gratefully, in particular from CERN and the ATLAS Tier-1 facilities at
TRIUMF (Canada), NDGF (Denmark, Norway, Sweden), CC-IN2P3 (France),
KIT/GridKA (Germany), INFN-CNAF (Italy), NL-T1 (Netherlands), PIC (Spain),
ASGC (Taiwan), RAL (UK) and BNL (USA) and in the Tier-2 facilities
worldwide.

\clearpage

\section{Expected number of events in MC and  90\% CL upper limits on the expected and observed number of signal events}


\begin{table}[ht!] 
\centering
\caption{The number of expected signal events for the eLJ--eLJ, muLJ--muLJ and eLJ--muLJ channels in $2 \gamma_d$ production. These numbers are obtained purely from simulations.}
  \begin{tabular}{m{29mm} |  >{\centering\arraybackslash}m{18mm} | >{\centering\arraybackslash}m{18mm} | >{\centering\arraybackslash}m{18mm} }
    \hline
Sample  (\ngammad = 2) & eLJ--eLJ & muLJ--muLJ & eLJ--muLJ \\
    \hline
\underline {SUSY MC} & & & \\
    \mgammad = 0.1 GeV  & 37.5 $\pm$ 1.0 & --  & -- \\
    \mgammad = 0.3 GeV  & 15.8 $\pm$ 0.6  & 14.7 $\pm$ 0.6  & 18.4 $\pm$ 0.7 \\
    \mgammad = 0.5 GeV  & 10.0 $\pm$ 0.5  & 14.7 $\pm$ 0.7 & 15.3 $\pm$ 0.6 \\
    \mgammad = 0.9 GeV  & 3.6 $\pm$ 0.2  & 9.3 $\pm$ 0.4 & 5.6 $\pm$ 0.3 \\
    \mgammad = 1.2 GeV & 6.3 $\pm$ 0.3 & 17.6 $\pm$ 0.5 & 10.2 $\pm$ 0.4 \\ 
    \mgammad = 1.5 GeV & 3.1 $\pm$ 0.2 & 9.6 $\pm$ 0.4 & 5.9 $\pm$ 0.3  \\
    \mgammad = 2.0 GeV & 3.6 $\pm$ 0.2 & 12.5 $\pm$ 0.5 & 6.8 $\pm$ 0.3 \\
    \hline
    \underline {Higgs MC} & & & \\
    \mgammad = 0.4 GeV with $10\%$ BR & 18.4 $\pm$ 1.2 & 103.7 $\pm$ 3.1 & 32.2 $\pm$ 1.7 \\
    \hline
     \end{tabular}
\label{tab:sig1}
\end{table}

\begin{table}[ht!] 
\centering
\caption{The number of expected signal events for all six LJ channels  in $4 \gamma_d$ production. These numbers are obtained purely from simulations.}
\begin{tabular}{m{22mm} |  >{\centering\arraybackslash}m{18mm} | >{\centering\arraybackslash}m{18mm} | >{\centering\arraybackslash}m{18mm} | >{\centering\arraybackslash}m{18mm} | >{\centering\arraybackslash}m{18mm} | >{\centering\arraybackslash}m{18mm} }
    \hline
Sample (\ngammad = 4)  & eLJ--eLJ & muLJ--muLJ & eLJ--muLJ & eLJ--emuLJ & muLJ--emuLJ & emuLJ--emuLJ \\
\hline
\underline {SUSY MC } & & & & & & \\
\mgammad = 0.1 GeV & 109.1 $\pm$ 1.6 & -- & -- & -- & -- & -- \\
\mgammad = 0.3 GeV & 12.0 $\pm$ 0.5 & 32.0 $\pm$ 0.9 & 21.5 $\pm$ 0.8  & 18.2 $\pm$ 0.7 & 30.0 $\pm$ 0.9 & 20.2 $\pm$ 0.8\\
\mgammad = 0.5 GeV & 4.5 $\pm$ 0.3 & 23.0 $\pm$ 0.8 & 11.2 $\pm$ 0.5 & 8.7 $\pm$ 0.5 & 17.3 $\pm$ 0.7 & 9.1 $\pm$ 0.5 \\
\mgammad = 0.9 GeV & 0.7 $\pm$ 0.1 & 8.4 $\pm$ 0.4  & 3.0 $\pm$ 0.2 & 2.0 $\pm$ 0.2  & 4.1 $\pm$ 0.3 & 2.4 $\pm$ 0.2 \\
\mgammad = 1.2 GeV & 2.1 $\pm$ 0.2 & 15.1 $\pm$ 0.5 & 6.8 $\pm$ 0.3 & 4.8 $\pm$ 0.3 & 11.6 $\pm$ 0.4 & 6.6 $\pm$ 0.3\\
\mgammad = 1.5 GeV & 0.8 $\pm$ 0.1 & 6.3 $\pm$ 0.3 & 3.1 $\pm$ 0.2 & 1.8 $\pm$ 0.2 &  3.3 $\pm$ 0.2 & 1.8 $\pm$ 0.2 \\
\mgammad = 2.0 GeV & 0.6 $\pm$ 0.1 & 5.3 $\pm$ 0.3 & 2.3 $\pm$ 0.2 & 1.7 $\pm$ 0.2 & 3.4 $\pm$ 0.2 & 1.8 $\pm$ 0.2 \\
    \hline
\underline {Higgs MC } & & & & &  \\
    \mgammad = 0.4 GeV with $10\%$ BR & 0.9 $\pm$ 0.6 & 23.5 $\pm$ 3.4 & 4.1 $\pm$ 1.4   & 3.3 $\pm$ 1.4 & 15.8 $\pm$ 2.7 & 4.0 $\pm$ 1.4 \\
\hline
    \end{tabular}
\label{tab:sig2}
\end{table}



\begin{table}[ht!]
{\centering \textbf{eLJ--eLJ} \par\medskip}
\footnotesize
\centering
\caption{This table shows the $95\%$ CL upper limit on the number of expected and observed events for various signal benchmarks in the {\bf{eLJ--eLJ}} channel. The standard deviations ($\sigma$) relates to the expected limit.}
\begin{tabular}{m{20mm} |  >{\centering\arraybackslash}m{7mm} | >{\centering\arraybackslash}m{7mm} | >{\centering\arraybackslash}m{7mm} | >{\centering\arraybackslash}m{7mm} | >{\centering\arraybackslash}m{7mm} |>{\centering\arraybackslash}m{7mm} || >{\centering\arraybackslash}m{7mm} | >{\centering\arraybackslash}m{7mm} | >{\centering\arraybackslash}m{7mm} |>{\centering\arraybackslash}m{7mm} |>{\centering\arraybackslash}m{7mm} |>{\centering\arraybackslash}m{7mm}}    \hline
    \hline
Signal benchmarks &  \multicolumn{6}{c||}{ $2 \gamma_d$ + X } & \multicolumn{6}{c}{ $2 (s_{d} \rightarrow \gamma_d \gamma_d)$ + X } \\
\cline{2-13}
$\gamma_d$ mass [GeV] &  Obs. & Exp.  & $-2 \sigma $ & $-1 \sigma$ & $+1\sigma$ & $+2 \sigma$ & Obs. & Exp. & $-2 \sigma $ & $-1 \sigma$ & $+1\sigma$ & $+2 \sigma$  \\
\hline
SUSY & & & & & & & & & & & &\\
\underline {($m_{\tilde{q}}= 700$ GeV)} & & & & & & & & & & & & \\

0.1&8.4&6.6&3.0&4.3&10.9&18.1&8.3&6.5&2.9&4.2&10.7&17.9  \\
0.2&8.4&6.6&3.0&4.3&10.9&18.1&8.3&6.5&2.9&4.2&10.7&17.9  \\
0.3&8.4&6.6&3.0&4.3&10.9&18.2&8.4&6.6&3.0&4.3&10.8&18.1  \\
0.4&8.4&6.6&3.0&4.3&10.9&18.3&9.0&7.1&3.2&4.6&11.7&19.5  \\
0.5&8.4&6.6&3.0&4.3&11.0&18.4&9.2&7.2&3.3&4.7&12.0&20.1  \\
0.7&8.5&6.7&3.0&4.3&11.0&18.6&9.3&7.3&3.3&4.7&12.1&20.4  \\
0.9&8.5&6.7&3.0&4.3&11.0&18.7&9.9&7.8&3.6&5.0&13.0&21.9  \\
1.2&8.6&6.7&3.0&4.3&11.2&18.9&9.3&7.3&3.3&4.7&12.1&20.5  \\
1.5&8.6&6.7&3.0&4.3&11.2&18.9&9.1&7.2&3.2&4.7&12.0&20.2  \\
2&8.5&6.7&3.0&4.3&11.1&18.8&9.1&7.2&3.2&4.7&12.0&20.3  \\

\hline
Higgs  & & & & & & & & & & & &  \\
\underline {($m_H = 125$ GeV)} & & & & & & & & & & & & \\
0.4 &8.5 &      6.7&    3.0&    4.3&    10.9&   18.3&   8.5&    6.7&    3.0&    4.3&    11.0&   18.3 \\

\hline
\hline
\end{tabular}
\label{tab:eLJeLJlimnum}
\end{table}

\begin{table}[ht!]
{\centering \textbf{muLJ--muLJ} \par\medskip}
\footnotesize
\centering
\caption{This table shows the $95\%$ CL upper limit on the number of expected and observed events for various signal benchmarks in the {\bf{muLJ--muLJ}} channel. The standard deviations ($\sigma$) relates to the expected limit.}
\begin{tabular}{m{20mm} |  >{\centering\arraybackslash}m{7mm} | >{\centering\arraybackslash}m{7mm} | >{\centering\arraybackslash}m{7mm} | >{\centering\arraybackslash}m{7mm} | >{\centering\arraybackslash}m{7mm} |>{\centering\arraybackslash}m{7mm} || >{\centering\arraybackslash}m{7mm} | >{\centering\arraybackslash}m{7mm} | >{\centering\arraybackslash}m{7mm} |>{\centering\arraybackslash}m{7mm} |>{\centering\arraybackslash}m{7mm} |>{\centering\arraybackslash}m{7mm}}    \hline
    \hline
Signal benchmarks &  \multicolumn{6}{c||}{ $2 \gamma_d$ + X } & \multicolumn{6}{c}{ $2 (s_{d} \rightarrow \gamma_d \gamma_d)$ + X } \\
\cline{2-13}
$\gamma_d$ mass [GeV] &  Obs. & Exp.  & $-2 \sigma $ & $-1 \sigma$ & $+1\sigma$ & $+2 \sigma$ & Obs. & Exp. & $-2 \sigma $ & $-1 \sigma$ & $+1\sigma$ & $+2 \sigma$  \\
\hline
SUSY & & & & & & & & & & & &\\
\underline {($m_{\tilde{q}}= 700$ GeV)} & & & & & & & & & & & & \\

0.3&5.9&6.5&2.9&4.1&10.9&18.7&6.3&6.9&3&4.4&11.6&19.9 \\
0.4&5.9&6.5&2.9&4.1&10.9&18.7&6.3&6.9&3&4.4&11.6&19.8 \\
0.5&5.9&6.5&2.9&4.1&10.9&18.7&6.4&6.9&3&4.4&11.7&19.9 \\
0.7&6&6.6&2.9&4.1&11&19&7&7.6&3.4&4.9&12.8&21.9 \\
0.9&5.9&6.5&2.9&4.1&10.9&18.7&6.7&7.3&3.3&4.7&12.3&21 \\
1.2&5.9&6.5&2.9&4.1&10.9&18.7&6.8&7.4&3.3&4.7&12.5&21.3 \\
1.5&5.9&6.5&2.9&4.1&10.9&18.7&7.1&7.8&3.5&4.9&13&22.1 \\
2&5.9&6.5&2.9&4.1&10.9&18.7&7.6&8.3&3.8&5.3&13.8&23.3 \\

\hline
Higgs  & & & & & & & & & & & &  \\
\underline {($m_H = 125$ GeV)} & & & & & & & & & & & &  \\
0.4&5.9&6.5&2.9&4.1&11&18.8& 6.5&7&3.1&4.5&11.9&20.3 \\

\hline
\hline
\end{tabular}
\label{tab:muLJmuLJlimnum}
\end{table}

\begin{table}[ht!]
{\centering \textbf{eLJ--muLJ} \par\medskip}
\footnotesize
\centering
\caption{This table shows the $95\%$ CL upper limit on the number of expected and observed events for various signal benchmarks in the {\bf{eLJ--muLJ}} channel. The standard deviations ($\sigma$) relates to the expected limit.}
\begin{tabular}{m{20mm} |  >{\centering\arraybackslash}m{7mm} | >{\centering\arraybackslash}m{7mm} | >{\centering\arraybackslash}m{7mm} | >{\centering\arraybackslash}m{7mm} | >{\centering\arraybackslash}m{7mm} |>{\centering\arraybackslash}m{7mm} || >{\centering\arraybackslash}m{7mm} | >{\centering\arraybackslash}m{7mm} | >{\centering\arraybackslash}m{7mm} |>{\centering\arraybackslash}m{7mm} |>{\centering\arraybackslash}m{7mm} |>{\centering\arraybackslash}m{7mm}}    \hline
    \hline
Signal benchmarks &  \multicolumn{6}{c||}{ $2 \gamma_d$ + X }& \multicolumn{6}{c}{ $2 (s_{d} \rightarrow \gamma_d \gamma_d)$ + X }  \\
\cline{2-13}
$\gamma_d$ mass [GeV] &  Obs. & Exp.  & $-2 \sigma $ & $-1 \sigma$ & $+1\sigma$ & $+2 \sigma$ & Obs. & Exp. & $-2 \sigma $ & $-1 \sigma$ & $+1\sigma$ & $+2 \sigma$  \\
\hline
SUSY  & & & & & & & & & & & & \\
\underline {($m_{\tilde{q}} = 700$ GeV)} & & & & & & & & & & & & \\

0.3&3.9&8&3.8&5.3&12.9&21& 4.1&8.6&4&5.7&13.8&22.3 \\
0.4&3.9&8&3.8&5.3&12.9&20.9& 4.1&8.6&4&5.7&13.8&22.4 \\
0.5&3.9&8&3.8&5.3&12.9&20.9& 4.3&8.8&4.1&5.8&14.1&22.8 \\
0.7&3.9&8.2&3.8&5.4&13.1&21.3 &5.3&10.7&5&7.1&16.9&27 \\
0.9&3.9&8.1&3.8&5.4&13&21.1& 5&10.2&4.8&6.8&16.1&25.8 \\
1.2&3.9&8.1&3.8&5.4&13&21& 4.7&9.7&4.6&6.5&15.4&24.7 \\
1.5&3.9&8.1&3.8&5.4&13&21& 5&10.1&4.8&6.8&16.1&25.7 \\
2&3.9&8.1&3.8&5.4&13&21.1& 5.6&11.1&5.4&7.4&17.4&27.5 \\

\hline
Higgs  & & & & & & & & & & & &  \\
\underline {($m_H = 125$ GeV)} & & & & & & & & & & & & \\
0.4&3.9&8.1&3.8&5.4&13&21.1& 4.5&9.2&4.4&6.1&14.7&23.8 \\

\hline
\hline
\end{tabular}
\label{tab:eLJmuLJlimnum}
\end{table}

\begin{table}[ht!]
{\centering \textbf{eLJ-- emuLJ} \par\medskip}
\footnotesize
\centering
\caption{This table shows the $95\%$ CL upper limit on the number of expected and observed events for various signal benchmarks in the {\bf{eLJ--emuLJ}} channel. The standard deviations ($\sigma$) relates to the expected limit.}
\begin{tabular}{m{20mm} |  >{\centering\arraybackslash}m{7mm} | >{\centering\arraybackslash}m{7mm} | >{\centering\arraybackslash}m{7mm} | >{\centering\arraybackslash}m{7mm} | >{\centering\arraybackslash}m{7mm} | >{\centering\arraybackslash}m{7mm}}    
\hline
    \hline
Signal benchmarks &  \multicolumn{6}{c}{ $2 (s_{d} \rightarrow \gamma_d \gamma_d)$ + X } \\
\cline{2-7}
$\gamma_d$ mass [GeV] &  Obs. & Exp.  & $-2 \sigma $ & $-1 \sigma$ & $+1\sigma$ & $+2 \sigma$  \\
\hline
SUSY  & & & & &  & \\
\underline {($m_{\tilde{q}} = 700$ GeV)} & & & & & &  \\

0.3&6.9&11.0&5.6&7.6&17.0&26.4  \\
0.4&7.3&11.7&5.9&8.0&17.8&27.7  \\
0.5&7.6&11.9&6.0&8.2&18.3&28.4  \\
0.7&6.5&4.5&2.3&3.0&7.0&10.7  \\
0.9&9.3&14.3&7.3&9.9&21.8&32.5  \\
1.2&10.1&14.9&7.8&10.4&22.2&32.7  \\
1.5&12.6&17.3&9.9&13.2&23.3&32.3  \\
2&13.7&17.4&10.7&13.5&23.1&32.0  \\

\hline
Higgs  & & & & & & \\
\underline {($m_H = 125$ GeV)} & & & & & &   \\
0.4&5.9&        9.7&    4.8&    6.6&    15.0&   23.6 \\
\hline
\hline
\end{tabular}
\label{tab:eLJemuLJlimnum}
\end{table}

\begin{table}[ht!]
{\centering \textbf{muLJ--emuLJ} \par\medskip}
\footnotesize
\centering
\caption{This table shows the $95\%$ CL upper limit on the number of expected and observed events for various signal benchmarks in the {\bf{muLJ--emuLJ}} channel. The standard deviations ($\sigma$) relates to the expected limit.}
\begin{tabular}{m{20mm} |  >{\centering\arraybackslash}m{7mm} | >{\centering\arraybackslash}m{7mm} | >{\centering\arraybackslash}m{7mm} | >{\centering\arraybackslash}m{7mm} | >{\centering\arraybackslash}m{7mm} | >{\centering\arraybackslash}m{7mm}}    \hline
    \hline
Signal benchmarks &  \multicolumn{6}{c}{ $2 (s_{d} \rightarrow \gamma_d \gamma_d)$ + X } \\
\cline{2-7}
$\gamma_d$ mass [GeV] &  Obs. & Exp.  & $-2 \sigma $ & $-1 \sigma$ & $+1\sigma$ & $+2 \sigma$  \\
\hline
SUSY & & & & & & \\
\underline {($m_{\tilde{q}} = 700$ GeV)} & &  & & & & \\

0.3&8.8&16.5&7.9&11&26.1&41.6 \\
0.4&8.8&16.5&7.9&11&26.2&42 \\
0.5&9&16.8&8.1&11.3&26.6&42.2 \\
0.7&26.1&42.4&19.8&28&54.6&74.4 \\
0.9&10.5&19.3&9.4&12.9&30.6&46.3 \\
1.2&9.3&17.1&8.4&11.6&26.9&42.3 \\
1.5&11.3&20.5&10.1&13.9&31.9&46.8 \\
2&11.2&20.2&10.1&13.8&31&45.6 \\

\hline
Higgs & & & & & &   \\
\underline {($m_H = 125$ GeV)} & & & & & &  \\

0.4&11.5&20.8&10.2&14&32.8&47.8 \\

\hline
\hline
\end{tabular}
\label{tab:muLJemuLJlimnum}
\end{table}

\begin{table}[ht!]
{\centering \textbf{emuLJ--emuLJ} \par\medskip}
\footnotesize
\centering
\caption{This table shows the $95\%$ CL upper limit on the number of expected and observed events for various signal benchmarks in the {\bf{emuLJ--emuLJ}} channel. The standard deviations ($\sigma$) relates to the expected limit.}
\begin{tabular}{m{20mm} |  >{\centering\arraybackslash}m{7mm} | >{\centering\arraybackslash}m{7mm} | >{\centering\arraybackslash}m{7mm} | >{\centering\arraybackslash}m{7mm} | >{\centering\arraybackslash}m{7mm} | >{\centering\arraybackslash}m{7mm}}    \hline
    \hline
Signal benchmarks &  \multicolumn{5}{c}{ $2 (s_{d} \rightarrow \gamma_d \gamma_d)$ + X } \\
\cline{2-7}
$\gamma_d$ mass [GeV] &  Obs. &  Exp.  & $-2 \sigma $ & $-1 \sigma$ & $+1\sigma$ & $+2 \sigma$  \\
\hline
SUSY  & & & & & & \\
\underline {($m_{\tilde{q}} = 700$ GeV)} & & & & & & \\

0.3&2.9&5.1&2.4&3.3&8.4&14.0  \\
0.4&2.9&5.1&2.4&3.3&8.4&14.0  \\
0.5&3.0&5.2&2.4&3.4&8.6&14.2  \\
0.7&3.1&5.4&2.5&3.5&8.8&14.7  \\
0.9&2.9&5.0&2.4&3.3&8.3&13.8  \\
1.2&2.9&5.1&2.4&3.4&8.5&14.1  \\
1.5&2.9&5.0&2.4&3.3&8.4&13.9  \\
2&3.0&5.3&2.5&3.5&8.7&14.4  \\

\hline
Higgs & & & & & &  \\
\underline {($m_H = 125$ GeV)} & & & & & & \\

0.4 &3.9&       6.1&    2.9&    4.0&    9.0&    13.3 \\

\hline
\hline
\end{tabular}
\label{tab:emuLJemuLJlimnum}
\end{table}


\clearpage
\bibliographystyle{atlasBibStyleWithTitle}
\bibliography{bibliography}
\clearpage
%

\begin{flushleft}
{\Large The ATLAS Collaboration}

\bigskip

G.~Aad$^{\rm 85}$,
B.~Abbott$^{\rm 113}$,
J.~Abdallah$^{\rm 151}$,
O.~Abdinov$^{\rm 11}$,
R.~Aben$^{\rm 107}$,
M.~Abolins$^{\rm 90}$,
O.S.~AbouZeid$^{\rm 158}$,
H.~Abramowicz$^{\rm 153}$,
H.~Abreu$^{\rm 152}$,
R.~Abreu$^{\rm 116}$
% ...
\bigskip
\\
% ...
$^{11}$ Institute of Physics, Azerbaijan Academy of Sciences, Baku, Azerbaijan\\
% ...
$^{85}$ CPPM, Aix-Marseille Universit{\'e} and CNRS/IN2P3, Marseille, France\\
% ...
$^{90}$ Department of Physics and Astronomy, Michigan State University, East Lansing MI, United States of America\\
% ...
$^{107}$ Nikhef National Institute for Subatomic Physics and University of Amsterdam, Amsterdam, Netherlands\\
% ...
$^{113}$ Homer L. Dodge Department of Physics and Astronomy, University of Oklahoma, Norman OK, United States of America\\
% ...
$^{116}$ Center for High Energy Physics, University of Oregon, Eugene OR, United States of America\\
% ...
$^{151}$ Institute of Physics, Academia Sinica, Taipei, Taiwan\\
$^{152}$ Department of Physics, Technion: Israel Institute of Technology, Haifa, Israel\\
$^{153}$ Raymond and Beverly Sackler School of Physics and Astronomy, Tel Aviv University, Tel Aviv, Israel\\
% ...
$^{158}$ Department of Physics, University of Toronto, Toronto ON, Canada\\
% ...
\end{flushleft}

\end{document}